\documentclass{aastex}          
\usepackage{spr-astr-addons}    

\begin{document}
%
\title{On the assessment of the nature of open star clusters and the determination of their basic parameters
with limited data}

\shorttitle{Open clusters' fundamental parameters}
\shortauthors{Carraro et al.}

   \author{Giovanni Carraro}
   \affil{Dipartimento di Fisica e Astronomia, Universit\`a degli Studi di Padova,
                    Vicolo dell'Osservatorio 3, I-35122, Padova, Italy.}
    \and
   \author{Gustavo Baume}
   \affil{Facultad de Ciencias Astron\'omicas y Geofísicas (UNLP), Instituto de Astrof\'isica
de La Plata (CONICET, UNLP), Paseo del Bosque s/n, La Plata, Argentina}
    \and
   \author{Anton F. Seleznev}
   \affil{Astronomical Observatory, Ural Federal University, Mira str. 19, Ekaterinburg, 620002, Russia}
  \and
  \author{Edgardo Costa}
   \affil{Departamento de Astronom\'ia, Universidad de Chile, Camino El Observatorio 1515, Santiago, Chile}

   \email{giovanni.carraro@unipd.it}

\begin{abstract}
Our knowledge of stellar evolution and of the structure and chemical evolution of the Galactic disk
largely builds on the study of open star clusters. Because of their crucial role in these relevant topics, large
homogeneous catalogues of open cluster parameters are highly desirable. Although efforts have been
made to develop automatic tools to analyse large numbers of clusters, the results obtained so far vary from
study to study, and sometimes are very contradictory when compared to dedicated studies of individual clusters. In this
work we highlight the common causes of these discrepancies for some open clusters, and show that at present dedicated studies yield a much
better assessment of the nature of star clusters, even in the absence of ideal data-sets.
We make use of deep, wide-field, multi-colour photometry to discuss the nature of
six strategically selected open star clusters: Trumpler~22, Lynga~6, Hogg~19, Hogg~21, Pismis~10 and Pismis~14.
We have precisely derived their basic parameters by means of a combination of star counts and
photometric diagrams. Trumpler~22 and Lynga~6 are included in our study because they are widely known, and
thus provided a check of our data and methodology. The remaining four clusters are very poorly known,
and their available parameters have been obtained using automatic tools only. Our results are in some cases in
severe disagreement with those from automatic surveys.
\end{abstract}


\section{Introduction}

In the last decade many attempts have been made to obtain basic parameters for large sets of open star
clusters (e.g. Kharchenko et al. 2005, 2013; Bukowiecki et al. 2011; Glushkova et al. 2010;
Loktin et al. 2001; Tadross 2011). The goals of these studies are obvious: first to provide large samples of
clusters with homogeneous fundamental parameters, and, second, eventually to use these samples to probe the
properties of the Galactic thin disk, where most open clusters reside, and/or of the open cluster
population as a whole. Some of these studies are limited to membership and parameter determination
(e.g. Kharchenko et al. 2013; Caetano et al. 2015), while other explore properties of the disk
(Popova \& Loktin 2005, 2008; Tadross 2014) or of the open cluster population as a whole
(Buckner \& Froebrich 2014; Kharchenko et al. 2016, and references therein).\

In general, these studies extract the necessary data from public surveys -- mostly the 2MASS (Skrutskie et
al. 2006) archive -- and use automatic star counts and quick inspection of photometric diagrams to decide
about the nature of a cluster, and then derive its parameters in a semi-automatic way, with different degrees
of sophistication (Krone-Martins \& Moitinho 2014, Caetano et al. 2015). The most popular of these catalogues
is that from Kharchenko al. (2005, 2013), which also provides proper motions, but the data are usually
limited to the brightest stars. A through-full and critical  comparison of the results coming out of these different data-sets has recently been
provided by Netopil et al. (2015).\\

Briefly, typical limitations of these works are: (1) the evaluation of  the clusters' reality  is based on just a
handful of stars, (2) there is a general lack of a proper error assessment (which is reflected by the artificially high
precision of the parameter estimates), (3) a systematic neglecting of previous investigations, leading
to gross mistakes in many cases, and (4) in many instances the plots used to infer clusters' properties
are very difficult to read. Large discrepancies
among different compilations are often found (Netopil et al. 2015), and one is  led to the
frustrating conclusion that the large number of star clusters in a compilation does not
compensate for the poor and quick analysis of the individual objects.
As we shall discuss here, a through-full data analysis is
unavoidable, and should focus on two crucial aspects: a careful visual inspection of the clusters surface density and photometric diagrams, and an exhaustive and critical literature search.\

It is well known that a truly complete analysis of a star cluster is extremely difficult: astrometric, photometric,
and spectroscopic data need to be acquired and analysed (see e.g. Curtis et al. 2013, Costa et al. 2015, and references therein).
In most cases only photometry is available. Nevertheless, if photometry covers the whole cluster
area, it is sufficiently deep, and it is multi-colour, it may be possible to reach solid conclusions about the physical
nature of a group of stars, by means of well-known, powerful, classical procedures. The results can obviously
be strengthened and refined later by incorporating spectroscopic and kinematic studies.\

In this work, we have selected a strategic sample of open clusters to illustrate how to obtain solid estimates
of the basic parameters of a star cluster based on limited data, and to highlight the common limitations of automatic surveys.
We present deep, wide field, multicolor
(\emph{UBVI$_{kc}$}) photometry and star counts for 6 clusters: Trumpler~22, Lynga~6, Hogg~19, Hogg~21, Pismis~10
and Pismis~14. Trumpler~22 and Lynga~6 are relatively well studied objects, and provided the means to check our
data and methodology. The former was included to illustrate an easy case, while the latter is clearly a more  complicated
one because of the large reddening value. We obtained new data for Hogg~19, for which only \emph{VI} data was available, and for which serious
discrepancies are present in the literature. We also present new data for Hogg~21, Pismis~10 and Pismis~14, for
which only semi-automatic parameter estimates are available from Kharchenko et al. (2013). In Table~1 we present
the coordinates of our targets.\\

\noindent
The layout of the paper is as follows. In Section~2 we present a careful scan of the available literature on
the clusters of interest, and  in Section~3 we discuss the observations and the reduction procedure; in Section~4 we
concisely describe our data analysis strategy, while in Section~5 we explain with some detail the methodology used for
the star counts. In Section~6 the full analysis of the various photometric diagrams is presented. Finally, in Section~7
we summarise our findings.

\section{Literature overview}

\noindent
{\bf Trumpler~22}

Lindoff(1968) obtained \emph{UBV} photographic photometry and concluded that the cluster is 100 Myr old and
is located at a distance of 1700 pc, for a reddening of E(B-V) = 0.51.
Previous distance estimates were based on a smaller number of stars, and resulted in larger distances:
1800 pc (Barkhatova 1950), 2210 pc (Trumpler 1930) and  2500 pc (Collinder 1931).
Similarly, estimates of the radius have resulted in a wide range of sizes: 7 arcmin (Trumpler 1930), 8 arcmin
Lindoff (1968), 10 arcmin (Collinder 1931) and 23 arcimn (Barkhatova 1950).\

The nature of Trumpler~22 as a physical star cluster has been questioned by Haug (1978). Based on photographic
material he concluded that Trumpler~22 is a spurious agglomeration. On the contrary, de La Fuente Marcos \&
de La Fuente Marcos (2009) suggested that Trumpler~22 forms a primordial pair with the nearby star cluster NGC~5617.
Recently, this latter possibility was scrutinised in detail by De Silva et al. (2015) by using CCD \emph{UBVI} photometry and
and high resolution spectroscopy. They concluded that Trumpler~22 is indeed a  real cluster, and that its fundamental
parameters (age, distance and
metallicity) do agree with those of NGC~5671, thus supporting the earlier suggestion that they form a binary cluster system.
Their proposed parameters are: an age of 70 Myr, a reddening E(B-V) =0.48 and a distance of 2100 pc.
For this cluster Kharchenko et al (2013) do provide a reddening consistent with other studies (0.50), but
obtained the smallest distance (1614 pc) and the oldest age (250 Myr).\\

\noindent
{\bf Lynga~6}

This is another example of a well studied object that has not always been considered a physical group. It is
famous mostly because it most probably hosts the Cepheid TW Normae. Moffat \& Vogt (1975) could not reach firm
conclusions about the cluster's nature because of their shallow photometry. Madore (1975) obtained a distance
of 2500 pc, for a reddening of E(B-V) = 1.35. The first CCD study is that from Walker (1985), who suggests that
the cluster is 100 Myr old, and is located at 2000 pc for a reddening of E(B-V) = 1.34.  Hoyle et al. (2003)
highlights the difficulties to determine the cluster parameters due to significant differential reddening.
This difficulty can be handled more easily in the infrared, as Majaess et al. (2011) demonstrated.  They
derived an age of 80 Myr and a distance of 1.9$\pm$0.1 kpc, for a reddening E(J-K) = 0.38$\pm$0.02.
For this cluster Kharchenko et al (2013) provide a reddening E(B-V) = 1.27, the smallest distance (1771 pc)
and the youngest age ($\sim$ 30 Myr).\\

\noindent
{\bf Hogg~19}

Seleznev et al. (2010) provide the only dedicated study available for this cluster. Their analysis of CCD
\emph{VI} and 2MASS \emph{JHK} photometry
suggests that Hogg~19 might be as old as 2.5 Gyr, and at a distance of 2.6 kpc, for a
reddening of E(B-V) = 0.65.  Bukowiecki et al. (2011) using 2MASS photometry suggest an age of 1.3 Gyr,
a reddening E(B-V)=0.60, and a distance of 3266 pc.
Kharchenko et al. (2013), using the vary same 2MASS data obtained different values for the basic parameters:
a distance of 898 pc, a reddening E(B-V) = 0.416, and an age slightly above 1.0 Gyr. The two studies rely on
very different approaches to derive the cluster parameters, and therefore Hogg~19 stands out as a particularly
crucial test-case for the present study.\\

\noindent
{\bf Hogg~21}

The only parameter determination available for this cluster is that from the large survey of Kharchenko et al.
(2013), who suggest the following solution: an age of 28.1 Myr, a distance of 1750 pc, and a reddening
E(B-V) = 0.729.\\

\noindent
{\bf Pismis~10}

The only parameter determination available for this cluster is that from the large survey of Kharchenko et al.
(2013), who suggest the following
solution: an age of 251.2 Myr, a distance of 8835 pc, and a reddening E(B-V) = 1.457\\

\noindent
{\bf Pismis~14}

The only parameter determinations available for this cluster are those from the large survey of Kharchenko et al.
(2013), who suggest the following
solution: an age of 223.8 Myr, a distance of 1775 pc, and a reddening E(B-V) = 0.479, and from
Bukowiecki et al. (2011), who suggest the following solution: 22 Myr for the age, 0.56 for  E(B-V) , and
a distance of 1314 pc.\\

\section{Observations and data reduction}

The observations were carried out with the Y4KCAM camera attached to the Cerro Tololo Inter-American
Observatory (CTIO, Chile) 1-m telescope, operated by the SMARTS consortium
\footnote{\tt http://http://www.astro.yale.edu/smarts}.
This camera is equipped with an STA~$4064\times4064$ CCD detector,
\footnote{\texttt{http://www.astronomy.ohio-state.edu/Y4KCam/detector.html}}
with 15-$\mu$m pixels, yielding a scale of 0.289$^{\prime\prime}$/pixel and a
field-of-view (FOV) of $20^{\prime} \times 20^{\prime}$ at the Cassegrain focus of the telescope.
This FOV is large enough to cover the whole clusters and to sample the surrounding Galactic field.
This is shown in Fig.~1 where we provide  V-band CCD images for the six fields.\\

\noindent
All observations were carried out in photometric, good-seeing conditions. Our \emph{UBVI} instrumental
photometric system was defined by the use of a standard broad-band Kitt Peak \emph{UBVI$_{kc}$} set of filters.
\footnote{\texttt{http://www.astronomy.ohio-state.edu/Y4KCam/filters.html}}
To determine the transformation from our instrumental system to the standard Johnson-Kron-Cousins system,
and to correct for extinction, each night we observed Landolt's areas PG~1047 and SA~98 (Landolt 1992)
multiple times, and with a wide range of air-masses.
Field SA~98 in particular includes over 40 well-observed standard stars, with a good magnitude and color
coverage: $9.5\leq V \leq15.8$, $-0.2\leq(B-V)\leq2.2$, $-0.3\leq(U-B)\leq2.1$. In Table~2 we present the
log of our observations.

\noindent
Basic calibration of the CCD frames was done using the Yale/SMARTS {\it y4k} reduction script based on the IRAF
\footnote{IRAF is distributed by the National Optical Astronomy Observatory, which is operated by the
Association of Universities for Research in Astronomy, Inc., under cooperative agreement with
the National Science Foundation.} package \textsc{ccdred}, and the photometry was performed using
IRAF's \textsc{daophot} and \textsc{photcal} packages. Instrumental magnitudes were extracted following
the point spread function (PSF) method (Stetson 1987) using a quadratic, spatially variable
master PSF (gaussian function). Finally, the PSF photometry was aperture corrected using
25 bright, not saturated, isolated stars across the whole field.\\

\noindent
The aperture photometry was carried out using the PHOTCAL package and we used transformation
equations of the form:\\

\noindent
u = U + u1 + u2 $\times$(U-B) + u3 $\times$X                         (1)\\
b = B + b1 + b2 $\times$(B-V) + b3 $\times$ X                        (2)\\
v = V + v1$_{bv}$ + v2$_{bv}$ $\times$ (B-V) + v3$_{bv}$ $\times$X   (3)\\
v = V + v1$_{vi}$ + v2$_{vi}$ $\times$ (V-I) + v3$_{vi}$ $\times$X   (4)\\
i = I$_{kc}$ + i1 + i2 $\times$ (V-I) + i3 $\times$ X                (5)\\

\noindent
where UBVI$_{kc}$ and $ubvi$ are standard and instrumental magnitudes respectively, and
X is the airmass of the observation. Subscripts 1 clearly refer to zero points, and 2 to color terms.
We adopted as mean values for the
extinction coefficients (subscripts 3) the typical values of  the CTIO site (see Baume et al. 2009). \\

\noindent
To derive V magnitudes, we used Eq.~3 when the B magnitude was available; otherwise we used Eq.~4.
The calibration coefficients and their uncertainties are shown in Table~3. In Fig.~2 the reader can
appreciate the trend of global (PSF plus calibration) photometric errors (specifically for Trumpler~22),
and notice that they are well below 0.05 mag down to V$\sim$19.5 for all color combinations.

\noindent
World Coordinate System (WCS) header information of each frame was obtained using the ALADIN tool
and 2MASS data (Skrutskie et al. 2006). This procedure allows us to obtain a reliable astrometric calibration ($\sim 0.12^{\prime\prime})$
and is explained in full details in Baume et al. (2009).

\noindent
We used the STILTS tool to manipulate tables and cross-correlate our $UBVI_{kc}$ with the $JHK$ 2MASS
data. We thus obtained a catalogue with astrometric/photometric information for all detected objects in
a FOV of approximately $20^{\prime} \times 20^{\prime}$ of each cluster region (see Fig.~1).
The full catalogues are made available in electronic form at the CDS website.\\

\noindent
As a sanity check of our photometry, we compared our data for Trumpler~22 with that of
De Silva et al. (2015), which was obtained with a different telescope and CCD detector.
From a grand-total of 1572 stars in common we obtained, in the sense of ours minus theirs:\\

\noindent
$\Delta$ V       = 0.03$\pm$0.08,\\
\noindent
$\Delta$ (B-V) =  0.03$\pm$0.07,\\
\noindent
$\Delta$ (U-B) = -0.06$\pm$0.11, and\\
\noindent
$\Delta$ (V-I)   = 0.02$\pm$0.10.\\

\noindent
Fig.~3 illustrates the nice agreement between the two data sets.

\section{Methodology}

We briefly summarise our methodology here. The interested readers can found a
more detailed description in Carraro \& Seleznev (2012).\

By definition, a Galactic cluster is a density enhancement above the general Galactic field. Therefore, the
first step is to define this over-density using star counts and estimate its radius. To this end, we
follow Seleznev (2016), where the procedure is fully described. It should be noted however, that
an over-density is only a good indication of the possible existence of a cluster, but not a proof of the
reality of a physical entity. In the Galactic disk, in many occasions over-densities are generated by random
fluctuations in extinction across the field of view and by chance alignments (Carraro 2006, and references
therein). Therefore, a close inspection of photometric diagrams -the classical two color diagram (TCD) and colour-magnitude
diagram- (CMD), is mandatory to check whether the stars generating the over density also produce well
defined photometric sequences.\

For this latter purpose we make extensive use of the \emph{U} filter, and apply the Q-method, as described
in Hiltner \& Johnson (1956). This method allows  us to identify stars sharing common reddening properties, and
derive their individual reddening, via comparison with a zero reddening, zero age main sequence (ZAMS).
In this study we adopt the ZAMS defined originally by Turner (1976, 1979), and later validated by
Turner \& Burke (2002) and Turner (2010).

Once corrected for reddening, stars are plotted in the reddening corrected CMD, where their distance is
estimated using the very same ZAMS, which is displaced only vertically to fit the star distribution.
This fit is normally done both in the $V_o/(B-V)_o$ and $V_o/(U-B)_o$ diagram, to ensure a simultaneous
solid fit. Finally, age is derived on the same diagrams employing solar metallicity isochrones from the Padova suite of
models (Bressan et al. 2012), that we adopt here for consistency with our previous studies (see, e.g.
Carraro \& Seleznev 2012). This latter study also describes how the errors of the basic parameters are
estimated.

In this study we refrain from complementing our photometric material with astrometric data. Any attempt we made
resulted in confused vector point diagrams. We ascribe this negative result to the crowding of our fields, and to the presence in 
many cases of high and patchy extinction.

\section{Star counts}

The surface density maps for our program clusters (see Fig.~4) were derived with the use of a kernel
estimator (Silverman 1986). We have used this method several times in the past (see e.g. Carraro \&
Seleznev 2012; Carraro et al. 2016), and we refer the reader to Seleznev (2016) for an exhaustive
description of it. These maps were derived using a kernel halfwidth ($h$) of 3
arcmin. Different density values are shown with different shades in Fig.~4. Their numerical values
are indicated in the shade scales. Axes show the distance from the center of the field in arcmin.
The positive direction of the Y axis coincides with
the direction to the North, while the positive direction of the X axis coincides with the direction to
the East. Star counts were also limited
to an area one $h$ away from the detector borders, to avoid under-sampling effects.\\

Together with other basic cluster parameters, our results on the clusters geometry are given in Table~4.
Columns 3 and 4 give the center coordinates determined by us: RA and DEC for 2000.0 equinox, in degrees,
and column 5 gives our estimated cluster radius in arcmin. The second column in the table shows the $V$
limiting magnitude ($V_{lim}$) that was adopted in each case to construct the surface density map.
This limiting magnitude was determined analysing the density maps at varying $V_{lim}$ in combination with a visual
inspection of the CMD of each cluster.

Each cluster centre was determined as a coordinates of maximum points of the corresponding linear densities
also plotted by the kernel estimator method (Seleznev et al. 2017). The centre of an open cluster is not
a well-defined value. It depends on limiting magnitude, the kernel halfwidth, the spectral band. In the
present work the limiting magnitudes were used as indicated in Table~4. The kernel halfwidth for the
linear densitiy was selected taking into account the density profiles for sample clusters.
The general condition for the cluster centre coordinates was the density profile without a non-physical
minimum at the centre of the clusters.\\

Using the clusters center coordinates previously obtained, radial density profiles were then derived,
again using the kernel estimator method. These profiles are shown in Fig.~5. The vertical axis is
the surface density in units of  $arcmin^{-2}$. The horizontal axis represents the distance from the cluster
center in arcminutes. As before, profiles are limited to a region one $h$ away from the detector border (which
corresponds to about one arcmin). Density profiles are shown with a thick solid lines, and the areas depicting
the 2$\sigma$--width confidence intervals are shown with dotted lines. These intervals were obtained by
means of a smoothed bootstrap estimate method (Seleznev 2016).\

We have also made visual estimates of the background stellar density, which are indicated with thin solid lines.
These were determined as follows. \\

\noindent
If the density profile exhibits an approximately flat area above the horizontal axis limit, then the background
density line is drawn taking into account the approximate equality of the square of areas between this line
and the density profile above and below it. The cluster radius is then estimated as the abscissa of
of the intersection of the density profile and the background density line. The corresponding
uncertainty is then the distance from the abscissa of the point of intersection of the confidence interval
line with the background density line at the cluster radius location. This was the case for Hogg~21,
Lynga~6, Pismis~10, and Trumpler~22, suggesting that these clusters might have an extended halo, and thus the
FOV investigated might not be large enough. In such cases it is more conservative to consider these radii
as lower estimates.\\
If the density profile does not show a flat area and decreases up to the limit of the horizontal axis,
then only an upper limit of the background density value can be estimated. In these cases,
obviously, only a lower limit of the cluster size can be inferred.

\section{Photometric diagram analysis}

In this section we deal with the interpretation of photometric diagrams constructed taking into account our
results from the star counts. They are mostly CMDs, but in some cases we also employ TCDs in the
$B-V/U-B$ plane to support our conclusions.

\subsection{Trumpler~22}

This well studied cluster was included among our targets as a sanity check of our photometry, and to
ensure that our method is producing reliable and reproducible results.
This cluster stands out as a significant density enhancement above the general Galactic field (see Fig~4, bottom right panel), and we
estimate its radius as 6.4$\pm$0.5 arcmin (see Fig.~5 and Table~4). Its center is slightly off-set with
respect to nominal published coordinates (see Table~1). In Fig~6 we show our parameter solution. In the left
panel we show the reddening derivation, done using a ZAMS. It was found to be E(B-V) =0.48$\pm$0.03.
The uncertainty was obtained by displacing the ZAMS back and forth along the reddening vector direction
until a fit was not possible anymore. Using the
Q-method we corrected all stars for their individual reddening, and thus derived a reddening corrected CMD,
which is shown in the right panel (only for the brightest stars). The red solid line is the same ZAMS as in
the left panel, displaced vertically by (m-M)$_o$=11.4$\pm$0.2. This implies a distance of 1.9$^{+0.2}_{-0.1}$,
in excellent agreement with De Silva et al. (2015). The black solid isochrone, taken from Bressan et al.
(2012), is for an age of 70 Myr, and does provide a nice fit to the distribution of stars seen in the CMD.
Our age estimation is in agreement with De Silva et al. (2015) as well.

\subsection{Lynga~6}

Basic parameters available for this cluster are considered quite solid. The infrared study by Majaess
et al. (2011) dealt successfully with the uncertainties in reddening and distance that affected previous
optical CCD photometry (Hoyle et al 2003), produced by the high and variable extinction in the FOV towards
the cluster. Our star counts analysis highlights a significant over-density at the cluster position (see
Fig.~4 mid-left panel), and implies a radius of 5.5$\pm$0.3 arcmin for Lynga~6. Stars selected within this distance from
the cluster center were used to construct the CMDs shown in Fig.~7 for three different color combinations.
There is a main sequence (MS) significantly wide in color because of variable reddening. An experienced eye
can also identify an interesting feature, namely a bifurcation of the MS at
$V\sim16$ and $(B-V)\sim1.2$, $(V-I)\sim1.5$, and $(B-I)\sim2.5$, which further widens the upper MS. We
argue that Lynga~6 lies in the red side of the sequence, while the blue side is simply the continuation of
the field star sequence.\

In Fig.~8 we present our parameter solution for Lynga~6. In the upper left panel we show a TCD, where we
plot a ZAMS to fit the distribution of the reddest stars, that we recognised in the CMD. This
ZAMS has been displaced by $E(B-V)$ = 1.25. We then derived stars' individual reddening using the Q-method, and their
distribution is shown in the upper right panel. This distribution peaks at
$E(B-V)$ = 1.2$\pm$0.1, which lends further support to our previous guess of the cluster reality and
location in the CMD. In the lower panel we present a reddening corrected CMD, from which we have estimated
distance and age. Beyond any doubt, stars sharing a common reddening produce a distinctive feature at a
distance of 2.0$^{+0.1}_{-0.1}$ kpc, as implied by the vertical shift of (m-M)$_o$ = 11.5$\pm$0.1 needed
to fit the data with a ZAMS (red solid line). This is  in excellent agreement with the infrared study of
Majaess et al. (2011). The black isochrone (displaced by the same amount in absolute distance modulus) is for an age
of 79 Myr, again in excellent agreement with Majaess et al. (2011). For this cluster, the parameter
estimates provided by Kharchenko et al. (2013) are in agreement with ours, with the exception of their age, which is
half our value.

\subsection{Hogg~19}

CMDs for this cluster are presented in Fig.~9. In these diagrams, we only included stars inside the
cluster radius (see Table~4), and for which the uncertainty in the $V$ magnitude is smaller than 0.02 mag
(see Fig.~2). The only evident features  are a well populated MS, and a tilted clump of stars at V $\sim$15 stretched
in the direction of the reddening vector (see Carraro \& Costa 2009, or Baume et al 2009, for very similar
occurrences). At first glance we cannot recognise any  obvious turn off point (TO).\\

\noindent
The left and middle panels show CMDs in the $V/B-V$ and $V/V-I$ planes, respectively. The right panel shows
the same CMD as in the middle panel, but with two  solar metallicity isochrones (Bressan et al. 2012) superimposed; an improved parameter solution close to the one by Seleznev et al. (2010, in red), and the other for that of Kharchenko et al. (2013,
in blue).  To make the solutions comparable, we transformed Kharchenko et al. (2013) reddening from E(B-V) into E(V-I),
using  E(V-I) = 1.244$\times$ E(B-V).
The solution in red adopts  E(V-I)=1.0, which provides a better fit than in Seleznev et al. (2010), 
and relies on the fact that the red clump might indicate the existence of an old star
cluster, for which the main sequence turnoff (TO) would be located at $V\sim18.7$, $(B-V)\sim1.3$. This TO
seems to correspond to a thick blue MS, which appears on top of the general distribution of stars. This sequence
has a different shape than the sequence brighter than $V$=18.7, and at this magnitude level one can appreciate
a significant change in the density of stars.
The blue solution, on the other hand, is in general quite a
poor fit. It seems to rely on the existence of a {\it clump} of only two stars, and on the assumption that the
TO coincides roughly with the brightest stars. If correct, it implies that Hogg~19 is a twin of the nearby star cluster NGC~6134,
which has fundamental parameters very similar to the ones of this fit.\\

\noindent
Clearly, little more can be extracted from these CMDs alone. The two solutions can be seen as equally valid,
and therefore a closer scrutiny is necessary. We decided therefore to adopt a different strategy,
and make use of the $U$-band data. In Fig.~10 we present a classical TCD in the $B-V/U-B$ plane,
which is well known to be very effective to identify common reddening early spectral type stars, and to separate populations
with different reddening. The solid black line is ZAMS,
that we include for reference, together with the direction of the reddening, indicated by an arrow in the
top-right corner of the diagram. In spite of the scattering, two groups can clearly be identified: one with
very low reddening, composed of stars of spectral type A0 and later, and most probably located close to the
Sun; and a second group, affected by larger reddening (0.7$\pm$0.1) -and hence more distant-
with stars of spectral type as early as B5, which we fit with the blue ZAMS. We tentatively
associate this latter group  with the Carina-Sagittarius arm (see below). In this TCD we see no trace of a population
with $E(B-V)\sim0.4$. This evidence excludes Kharchenko et al. (2013) solution.
There is  no trace of stars as faint as the Hogg~19 MS which would support the suggestion
by Seleznev et al. (2010). This is because  the U photometry is not deep enough. In fact
the TO in this case is too shallow, and for its (B-V) color there are only very few stars
having also a measure in (U-B) (see Fig.~9 and 10).

\noindent
To further discriminate between the above two possibilities, we resorted to star counts and produced
surface density maps for the stars in the two magnitude regions that better highlight each solution (from the right panel of Fig.~9 ). The
results are shown in Fig.~11. The left panel represents stars brighter than $V$=14 (and E(B-V)$\sim$0.4, see above), which should produce
evidences supporting the suggestion by Kharchenko et al. (2013). Although there is a sort of elongated
concentration above and to the right, of the center of the field, the density contrast is so low that we
can safely conclude that the distribution of bright stars across the field is homogeneous. We suggest this sparse group
is made of interlopers, namely a few young stars located in the intervening Carina-Sagittarius spiral arm.

On the other
hand, the right panel shows the density map for stars in the magnitude range $18\leq V \leq20$, which would
belong to the TO of Hogg~19, according to Seleznev et al. (2010). In this case the density contrast is much
higher, and although the structure of the concentration is irregular, a peak is clearly visible.\\

These evidences altogether lend support Seleznev et al. (2010) suggestion, that we revised here . Kharchenko et al. (2013)
could not detect the cluster because of the shallowness of their photometry, and ended
up confusing it with a sparse group of young stars belonging to the intervening Carina Sagittarius arm  (see Carraro et al. 2010
for a similar concurrence).

\subsection{Hogg~21}

This cluster presents an easier case compared to Hogg 19. It stands as a low contrast over density in the top-right panel
of Fig.~4.
The CMD in Fig.~12 shows two distinct features: a thick sequence with a TO at V$\sim$ 17, and a
scattered group of blue stars brighter than $V\sim15$. For this latter group one can guess a TO at $V\sim13$.
In a TCD constructed with the stars within the cluster radius (see left panel of Fig.~13), a clear sequence of
young star appears, reddened by E(B-V) = 0.48$\pm$0.02, as indicated by the red ZAMS.  At odds with Hogg~19,
the sequence of young stars is much less scattered, and corresponds to a spatially confined structure,
enhancing the probability that we are facing a physical cluster and not a random distribution of young stars.
Using the Q method, we derived stars' individual reddening, and plotted the early type stars in the reddening-corrected
CMD shown in the right panel of Fig~13, from which a distance modulus of $(m-M)_o$=11.65$\pm$0.10 is implied.
Some scatter is clearly visible, but we ascribe it to the presence of binary stars, and to photometric errors,
particularly in the $(U-B)$ color index. This distance modulus places Hogg~21 at 2.1$^{+0.1}_{-0.1}$ kpc from
the Sun. The isochrone super-imposed in the figure is for an age of 100 Myr, which we inferred from the
earliest spectral type stars present in the cluster; i.e., about B6 (see also the TCD). Our estimates of the
cluster distance and age do not differ much from those of Kharchenko et al. (2013), but reddening does, meaning
that the apparent distance modulus from Kharchenko (2013) is clearly off. We note that significant systematic offsets
in the reddening estimates of Kharchenko et al. (2013) have been already reported in the literature (see, e.g.,
Buckner \& Froebrich 2014, 2016; Netopil et al. 2015).

\subsection{Pismis~10}

The surface density map for  the FOV in the direction to Pismis~10, shown in the mid-right panel of Fig.~4, indicates
a mild, wide, over density. However, a quick look at Fig.~1 (or any DSS map) suggests that this over density might be a reddening
effect, since the surrounding region is highly obscured. The CMDs presented in Fig.~14 hardly indicate any
distinctive feature in any of the color combinations, but the TCD for stars inside the cluster radius (see
Fig.~5 and Table~4) presented in the left panel in Fig.~15 shows a group of extremely and differentially reddened
stars, which might constitute an obscured star cluster, in full similarity with Lynga~6. The reddening solution
for this group is in fact $E(B-V)$=1.5$\pm$0.1. After correcting these stars for individual reddening, they distribute in
the reddening corrected CMD ( right panel of Fig.~15) in a cluster of $\sim$ 250 Myr, at a distance of less than 2.7$^{+0.3}_{-0.2}$ kpc
[(m-M)$_o$=12.2$\pm$0.2]. In this case, our age and reddening are close to Kharchenko et al. (2013).
Their distance estimate however differs by a very large amount. Again, this is caused by the insufficient depth of their photometry.

\subsection{Pismis~14}

On maps this cluster appears as a shallow over density a few arcmin from the more conspicuous open cluster NGC~2910 (Giorgi et al. 2015),
from which it is separated by a clear dust lane (see the bottom-left panels of Figs.~1 and 4, or any DSS image).
As such, Pismis~14 might simply constitute a random enhancement of a few bright stars,
or, possibly, be part of the outer corona of NGC~2910. The CMDs presented in Fig.~16 illustrate this latter
possibility. When considering stars outside the estimated radius of Pismis~14, a MS is clearly visible, which
would belong to NGC~2910 (right panel). On the contrary, the stars inside the cluster region (left panel)
do not show any distinctive feature.\

\noindent
The TCD for all stars in the FOV of Pismis~14 in Fig.~17 indicates a group of early type stars,
that must  belong to NGC~2910. They are reddened by $E(B-V)$ = 0.20, in perfect agreement with the recent
study of  NGC~2910 by Giorgi et al. (2015). These authors as well stress the large differences between their cluster
parameters and those determined by Kharchenko et al. (2013). Our conclusion is that Pismis~14 is a change alignment
of a few bright stars enhanced by a patchy reddening distribution. These stars are probable peripheral members of NGC~2910.

\section{Discussion and conclusions}

We have presented and discussed multi-color CCD photometry of six Galactic open clusters: Trumpler~22, Lynga~6,
Hogg~19, Hogg~21, Pismis~10, and Pismis~14. The main goal of the present study was to assess the nature of these
clusters and derive their basic parameters: reddening, age, and distance. Comparison of our results with the
literature has allowed us to highlight (and possibly explain) existing discrepancies with
large semi-automatic parameter surveys, which stresses their present limitations.\\

\noindent
Trumpler 22 and Lynga 6 are well studied stars clusters, and we have obtained results consistent with previous
dedicated works.  For Lynga~6, the extensive usage of U-band photometry allowed us
to reproduce infrared results  in a critical case of heavy reddening.

\noindent
Hogg~19 revealed itself as an extremely challenging object, and previous studies have been vastly discrepant
in their results (Kharchenko et al. 2013; Seleznev et al. 2010). A careful and synoptic analysis of both star counts and
photometric diagrams show this is  a $\sim$ 2 Gyr old cluster.
This is quite an interesting results, given the general paucity of old clusters in the inner Galaxy (Carraro et al. 2014,
Jacobson et al. 2016).  In this case the cause of the discrepancy is twofold:  first, the photometry used by Karchenko et al. (2013) is too shallow and they missed the real cluster; second, they did not take into account previous literature results.\\

\noindent
We found that Hogg~21 is a young star cluster associated with the Carina Sagittarius arm. It is significantly
less reddened than the estimate by Kharchenko et al. (2013). Systematic issues with their reddening estimates
have been routinely reported in the literature (Buckner \& Froebrich 2014; Netopil et al. 2015).\\

\noindent
Pismis 10 attracted our attention because of the very large heliocentric distance ($\sim$ 9 kpc) reported by
Kharchenko et al. (2013). This would place it well beyond the Vela Molecular Ridge (Carraro \& Costa 2010;
Giorgi et al. 2015) at the extreme periphery of the Galactic disk.
We found that this cluster is very reddened, as Lynga~6, but lies much closer, at  about 3 kpc from the Sun.\\

\noindent
We finally provide evidences that Pismis 14 is not a physical cluster, but a chance alignment of a few bright stars
member of the outer corona of the nearby cluster NGC~2910.\\

In closing, we acknowledge that efforts are being made by several groups (Caetano et al. 2015; Krone-Martins \& Moitinho 2014;  Netopil  et al. 2015; Carraro et al. 2016, and references
therein) to provide new tools, or re-discover old methods, to derive reliable open cluster parameters.
Hopefully, the closed box era of open clusters' fundamental parameter determination will be over soon.
Along this vein, future Gaia mission data releases will surely be very valuable to ensure the homogeneity of star cluster
fundamental parameters for all those clusters which will be at reach.

\acknowledgments
G. Baume acknowledges financial support from the ESO visitor program that allowed a visit to ESO premises
in Chile, where part of this work was done.
G. Carraro  science leave  in Ekaterinburg (where most of this was was done) was supported by Act 211 Government
of the Russian Federation, contract No. 02.A03.21.0006, and by the ESO DGDF program. E. Costa acknowledges
support by the Fondo Nacional de Investigaci\'on Cient\'ifica y Tecnol\'ogica (proyecto No. 1110100 Fondecyt)
and the Chilean Centro de Excelencia en Astrof\'isica y Tecnolog\'ias Afines (PFB 06).

{\it Facilities:} \facility{CTIO SMARTS}.

\clearpage

\begin{figure*}
\includegraphics[scale=0.7]{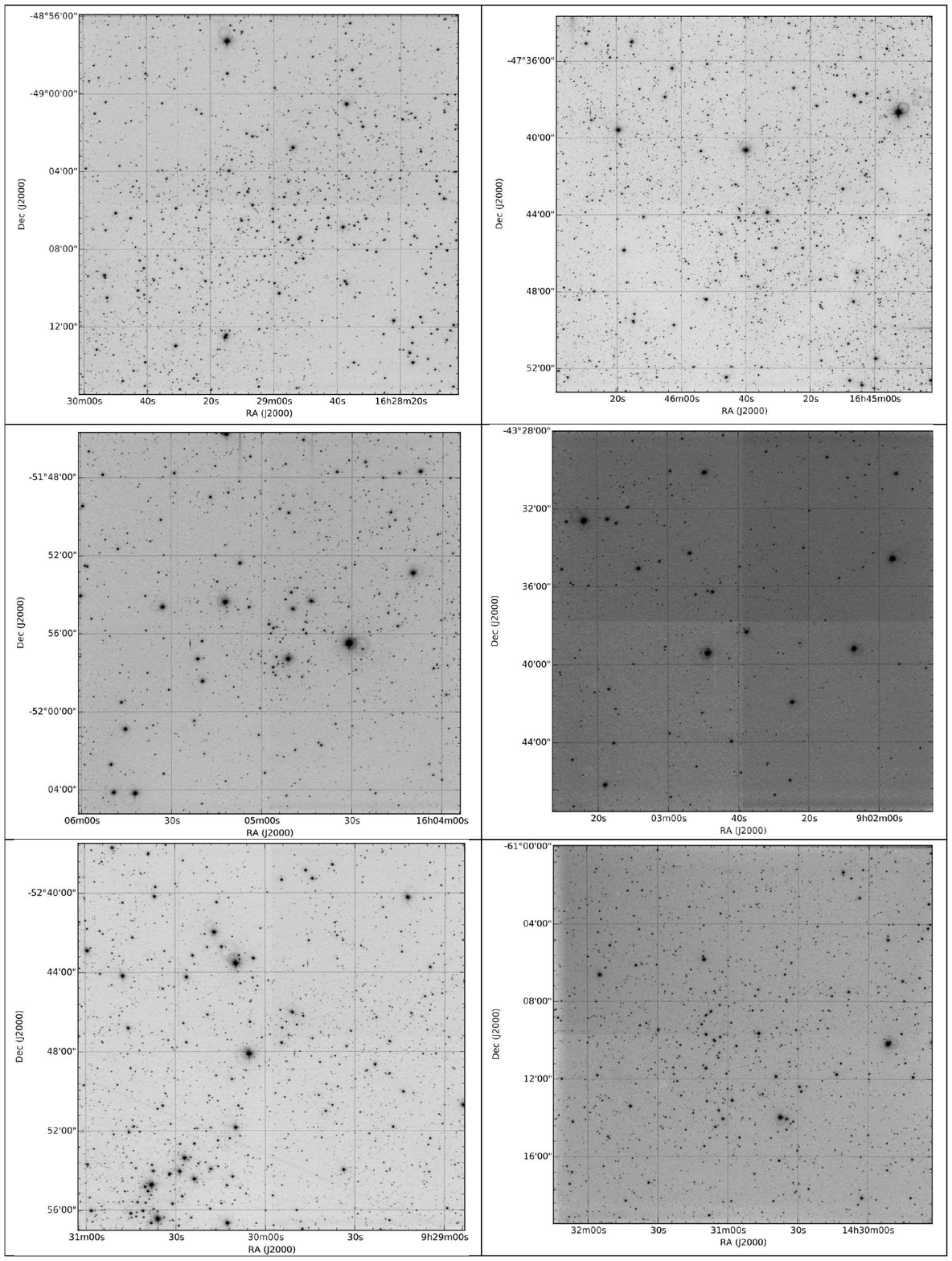}
\caption{$V$-band CCD images of the fields observed. Top row: Hogg~19 and 21.
Middle row: Lynga~6 and Pismis~10. Bottom row: Pismis~14 and Trumpler~22.}
\end{figure*}

\clearpage

\begin{figure}
\includegraphics[width=\columnwidth]{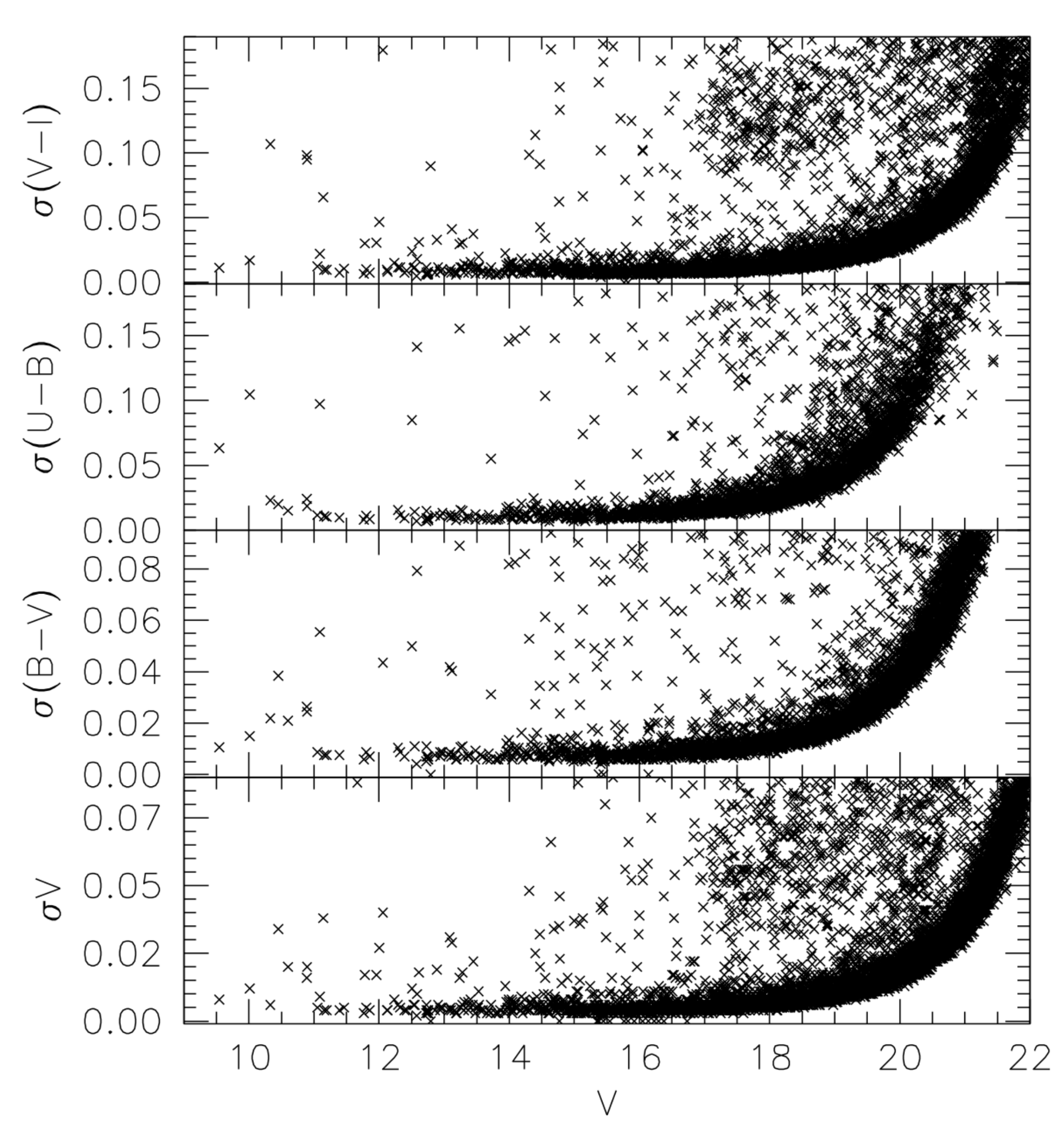}
\caption{Global photometric errors as a function of $V$ magnitude for Trumpler~22.}
\end{figure}

\clearpage

\begin{figure}
\includegraphics[width=\columnwidth]{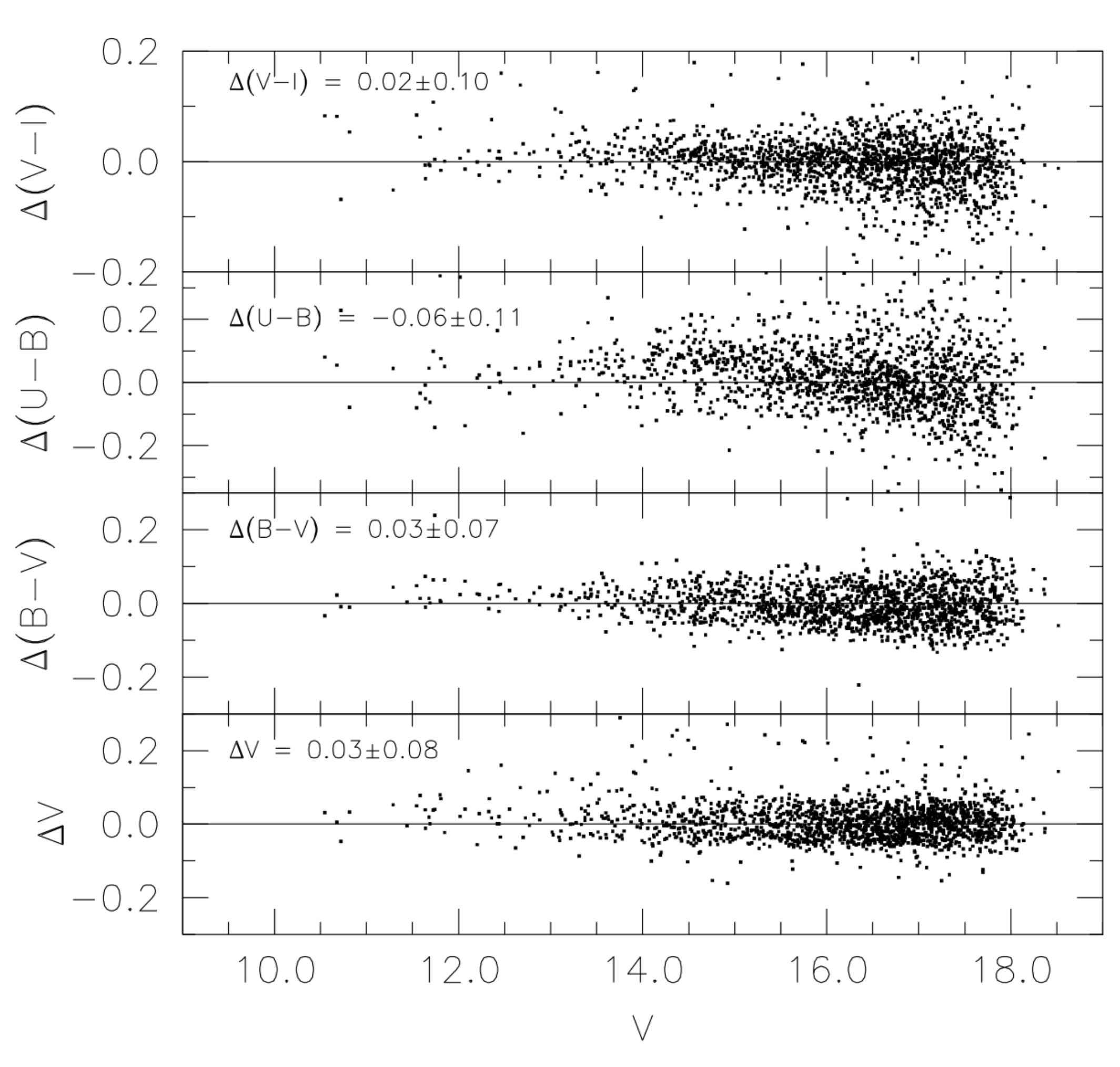}
\caption{Comparison of our photometry for Trumpler~22 with that of De Silva et al. (2015), in
the sense of our data minus theirs.}
\end{figure}

\clearpage

\begin{figure*}
\includegraphics[scale=0.75]{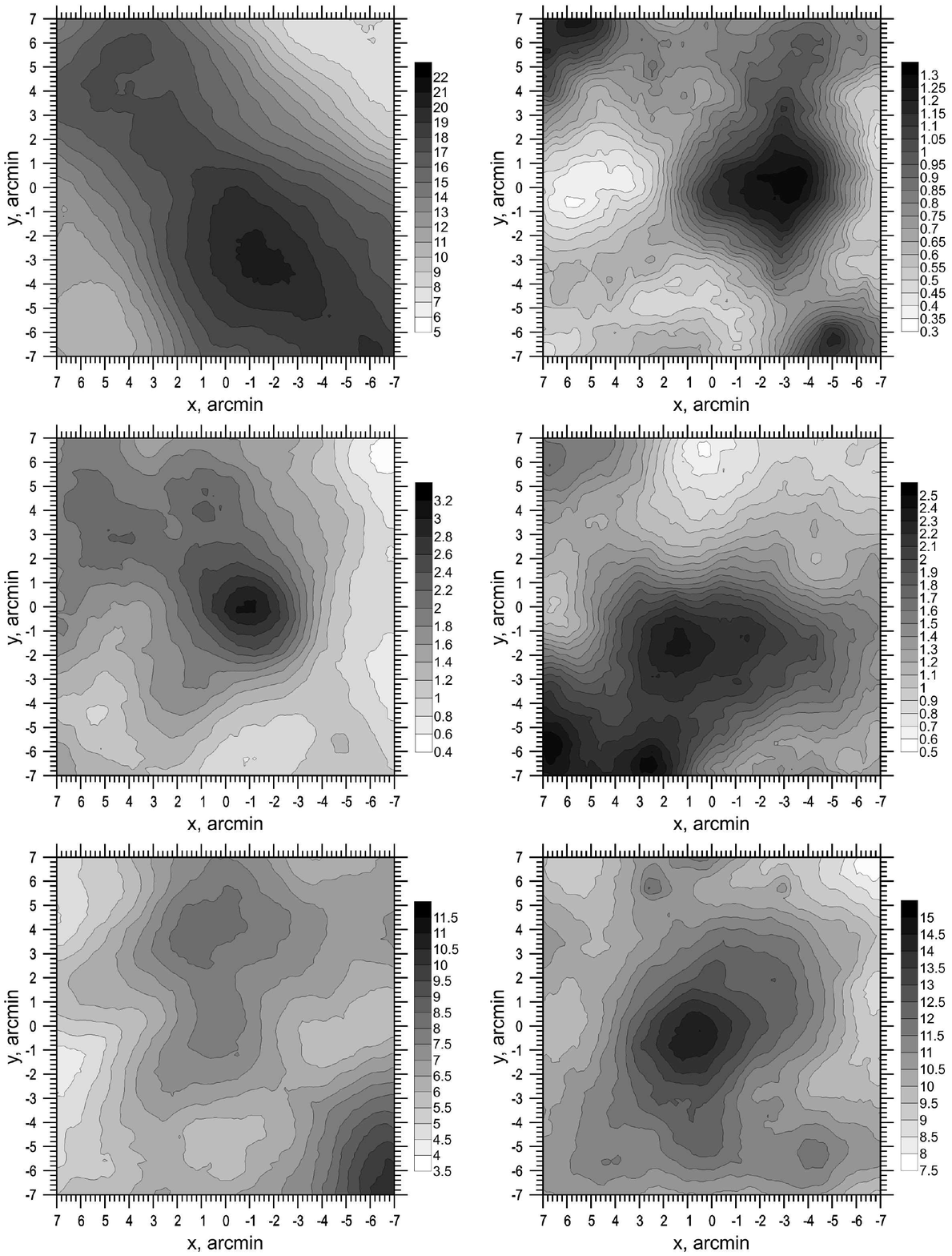}
\caption{Surface density maps. From the top to bottom: Hogg~19 and Hogg~21, Lynga~6 and Pismis~10,
Pismis~14 and Trumpler~22 .}
\end{figure*}

\clearpage

\begin{figure*}
\includegraphics[scale=0.75]{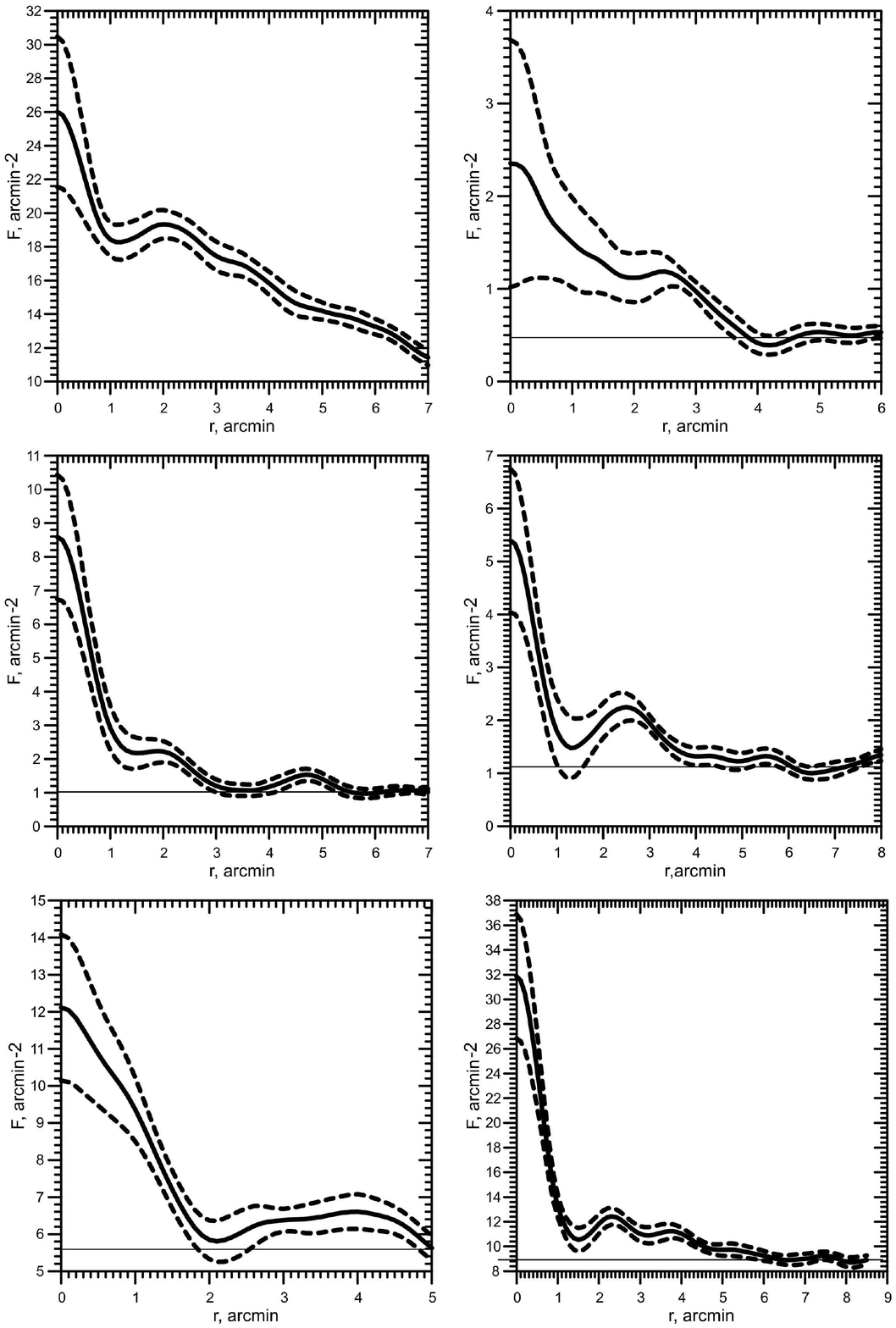}
\caption{Radial density profiles. From top to bottom: Hogg~19 and Hogg~21, Lynga~6 and Pismis~10,
Pismis~14 and Trumpler~22 .}
\end{figure*}

\begin{figure}
\plottwo{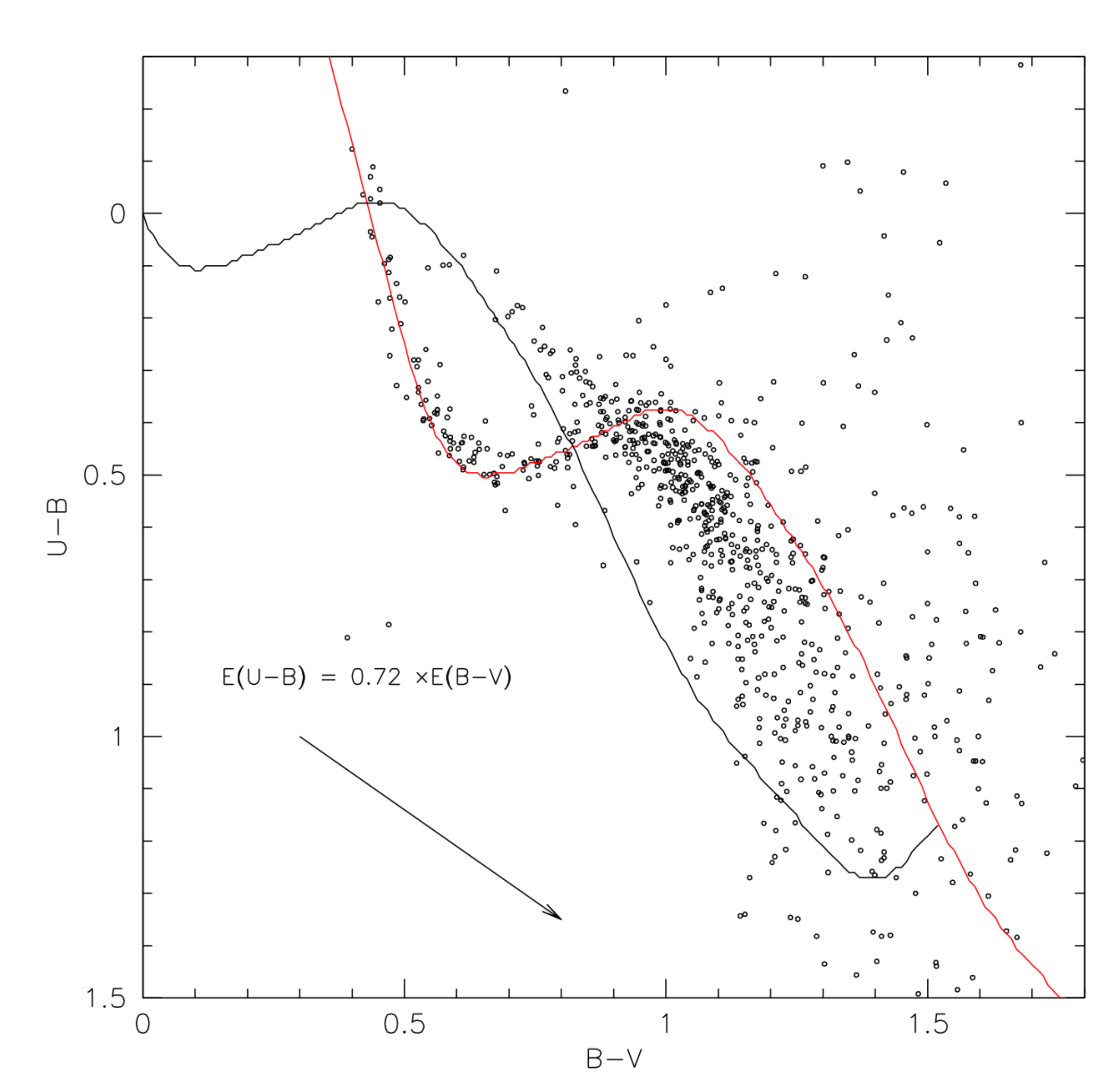}{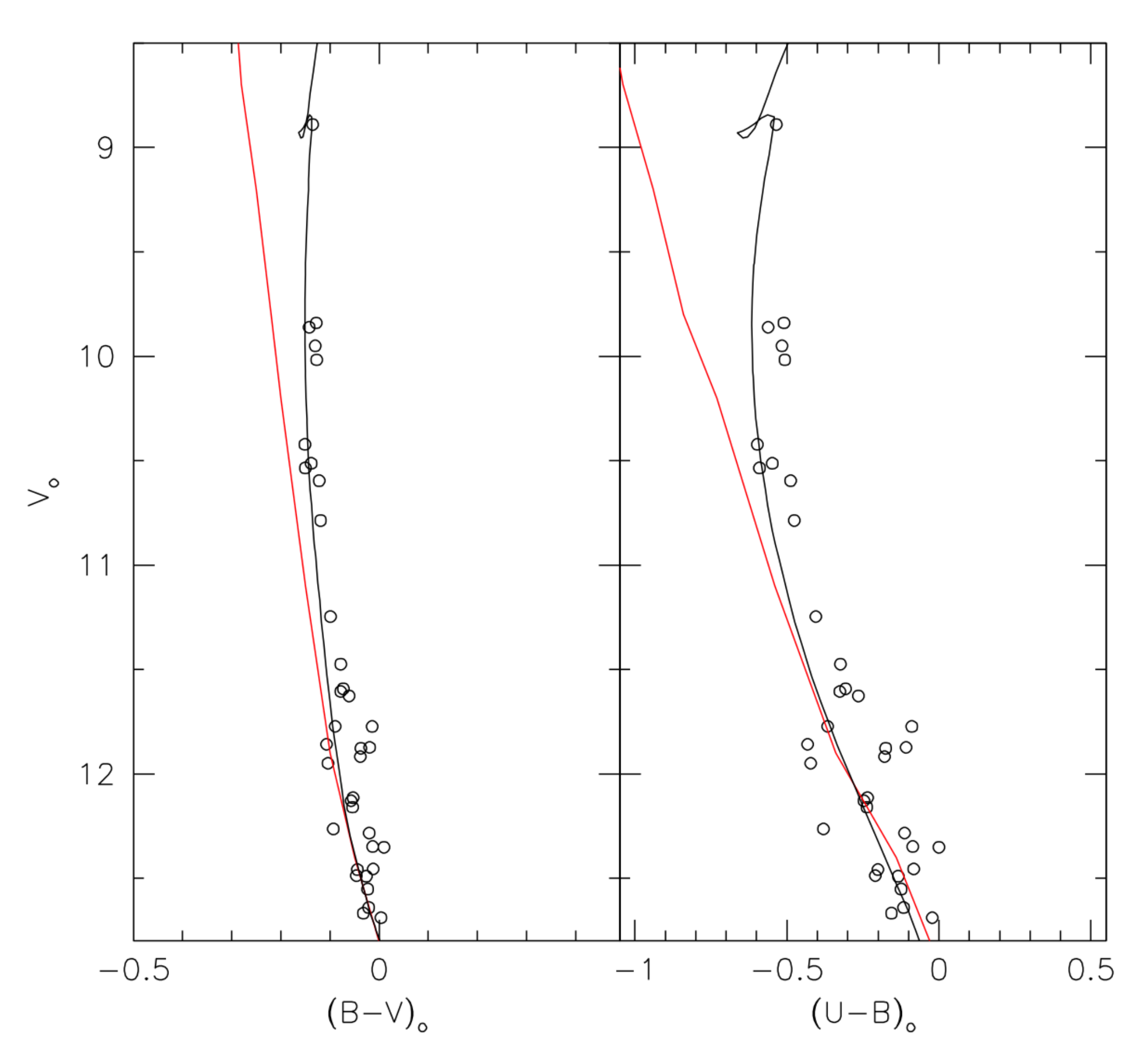}
\caption{Parameter solution for Trumpler~22. The left panel shows the TCD, and the reddening
solution. The solid black and red lines are ZAMS. The red ZAMS has been shifted by $E(B-V)=0.48$ in the
direction of the reddening vector. The right panel shows the reddening corrected CMD used for our distance
and age solution. The red line is a ZAMS displaced vertically by (m-M)$_o$=11.4; the black line is an
isochrone for an age of 70 Myr.}
\end{figure}

\begin{figure}
\includegraphics[width=\columnwidth]{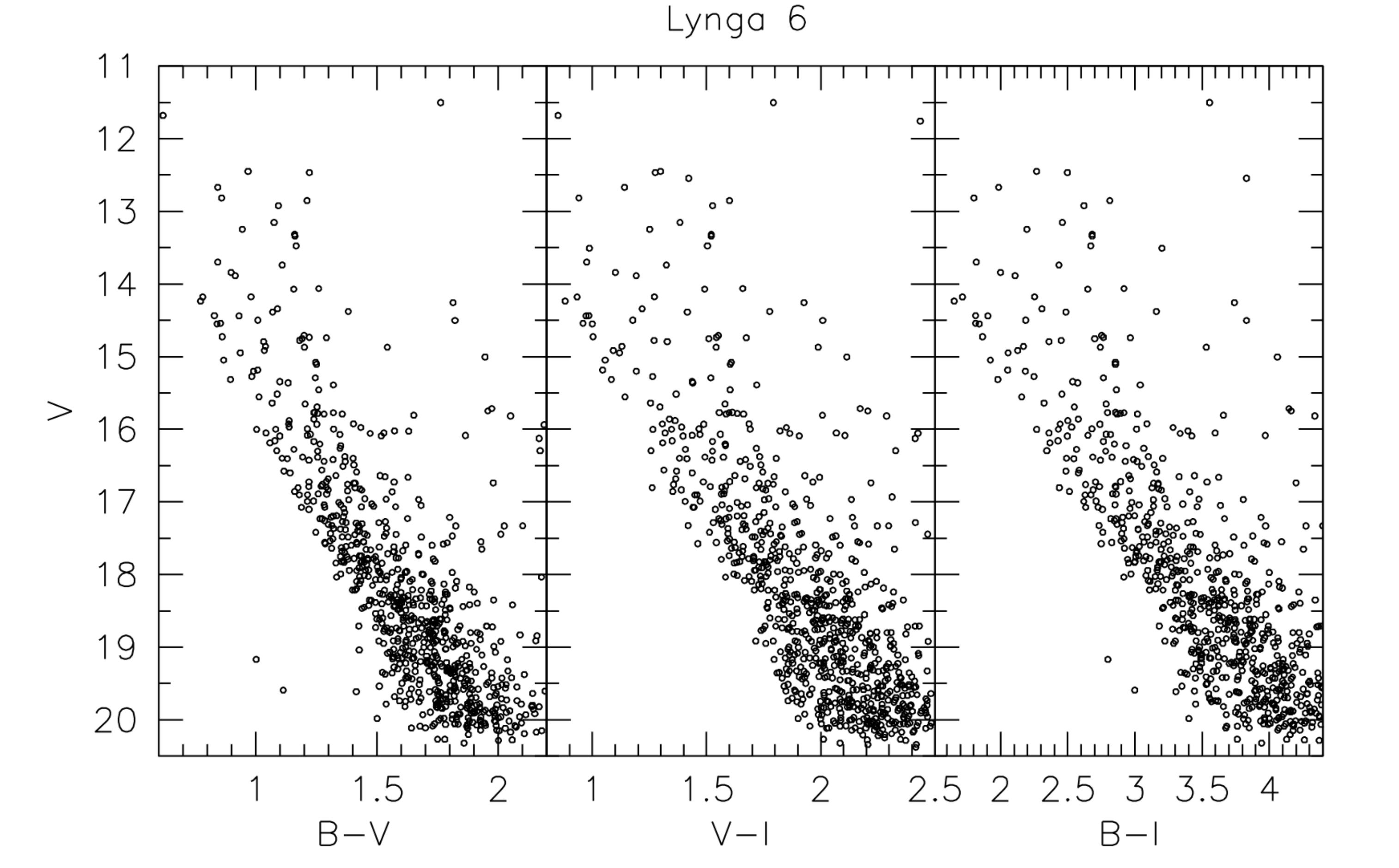}
\caption{CMDs, in three different color combinations, for Lynga~6 stars located inside the cluster radius.}
\end{figure}

\begin{figure*}
\includegraphics[scale=0.38]{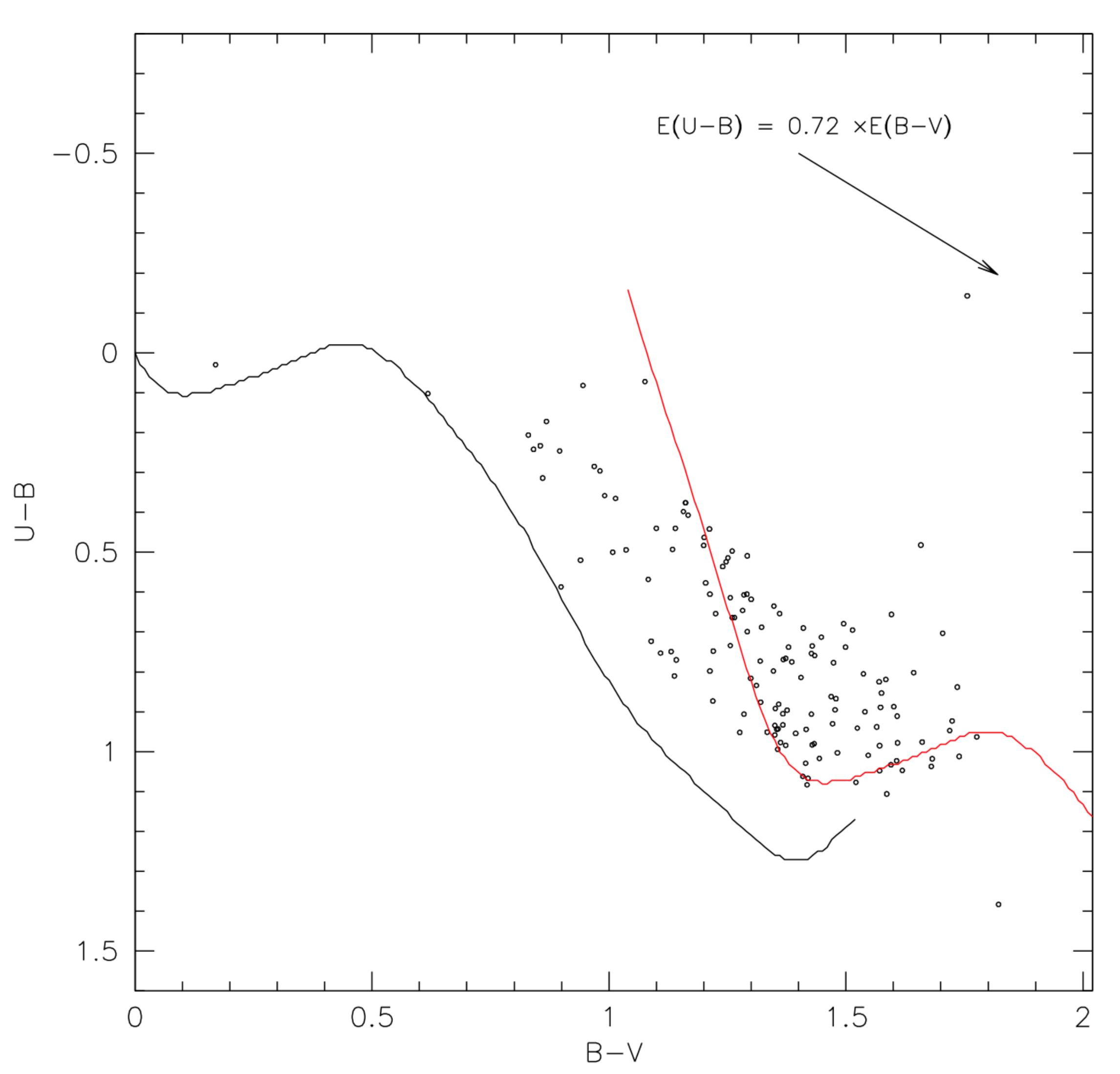}
\includegraphics[scale=0.38]{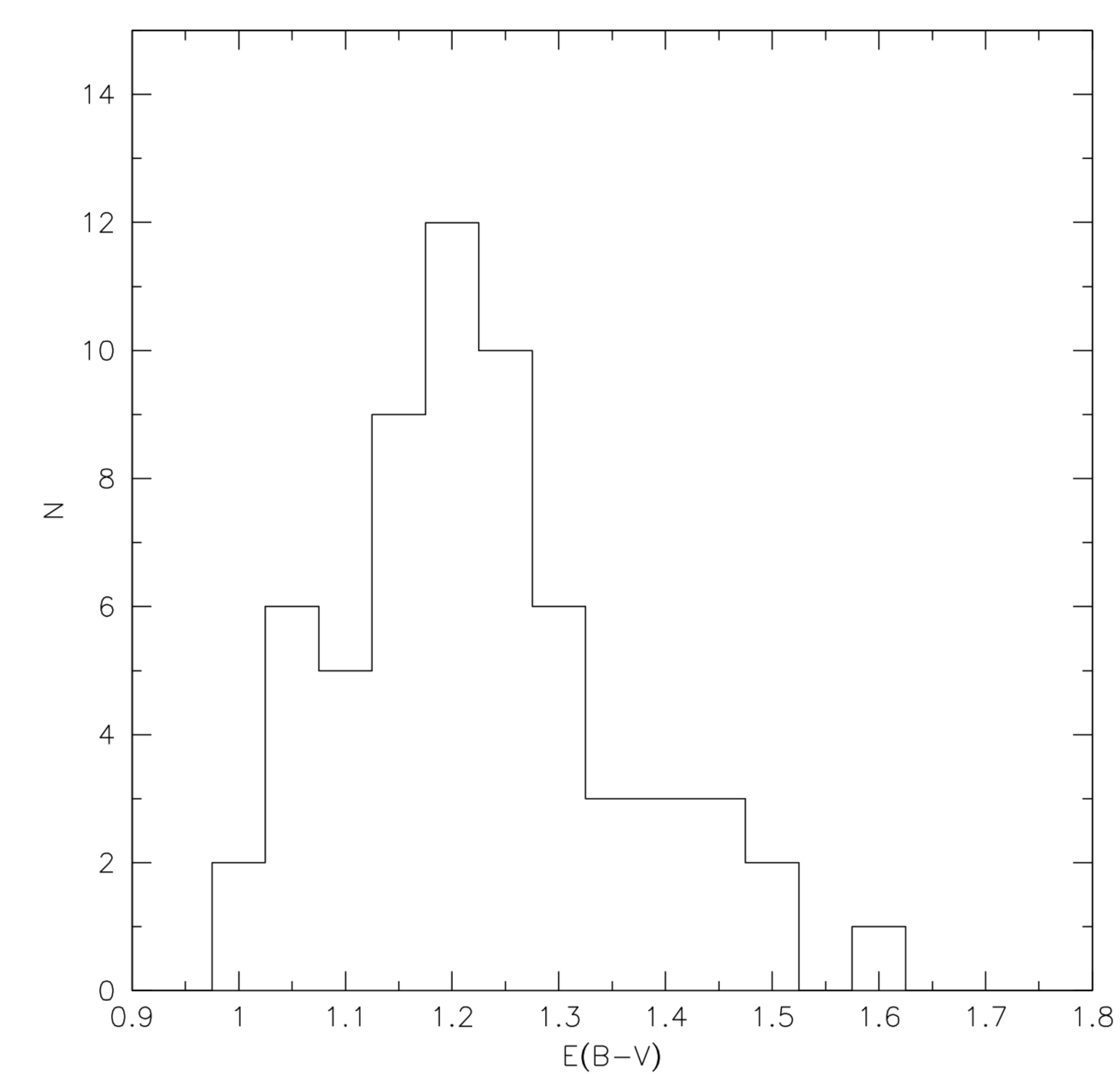}
\includegraphics[scale=0.4]{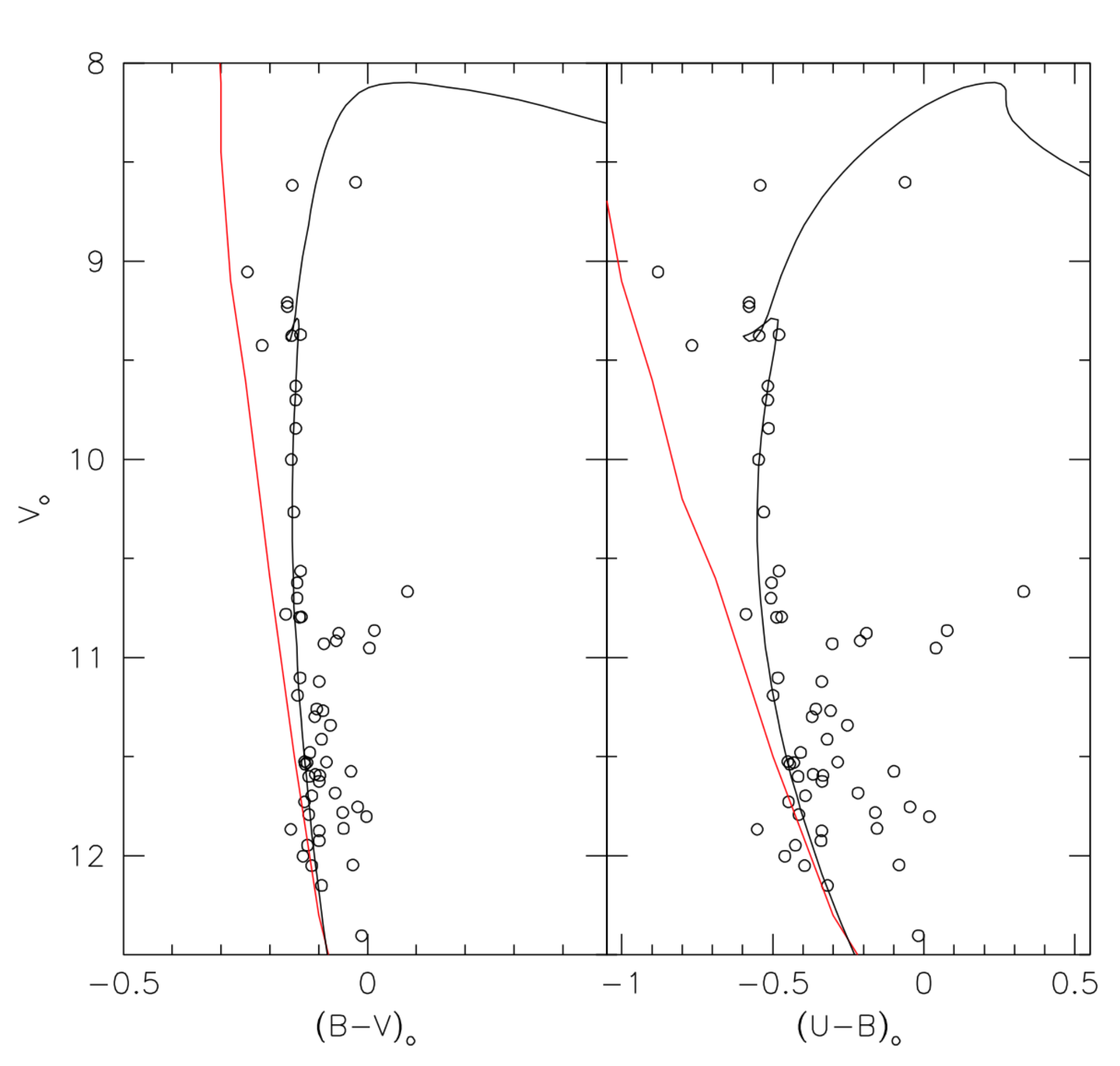}
\caption{Parameter solution for Lynga~6. The upper left panel shows the TCD, where the red line is a
ZAMS shifted by $E(B-V)=1.25$. The upper right panel illustrates the individual reddening distribution.
The lower panel, finally, shows the distance and age solution. The red line is a ZAMS
displaced vertically by (m-M)$_o$=11.5, and the black line is an isochrone for an age of 79 Myr.}
\end{figure*}

\begin{figure}
\includegraphics[width=\columnwidth]{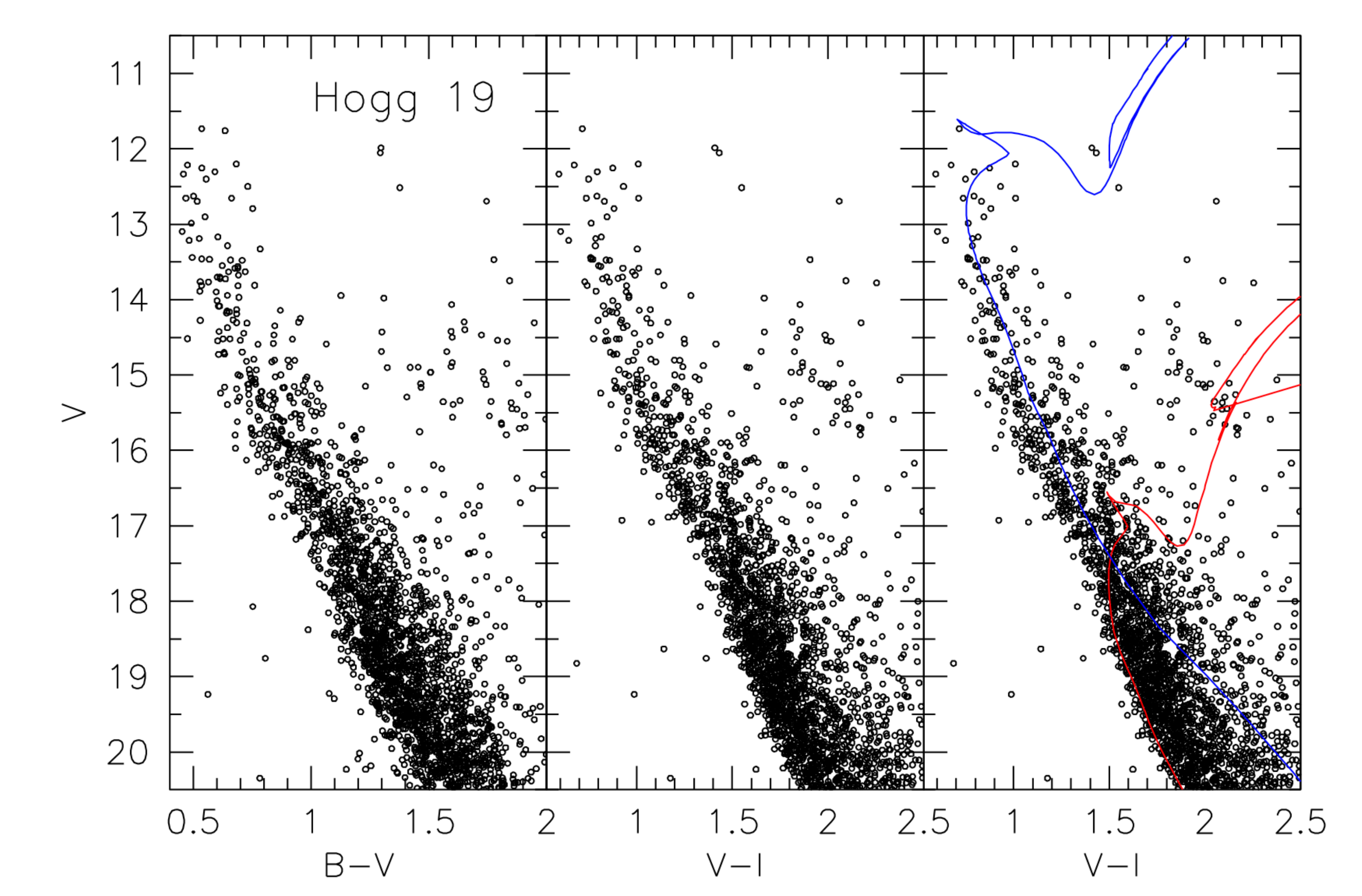}
\caption{CMDs of Hogg~19 in $V/B-V$ (right panel) and $V/V-I$ (middle panel). In the right panel we included
the parameter solutions from Kharchenko et al. (2013, in blue) and the one from this study (in red).}
\end{figure}

\begin{figure}
\includegraphics[width=\columnwidth]{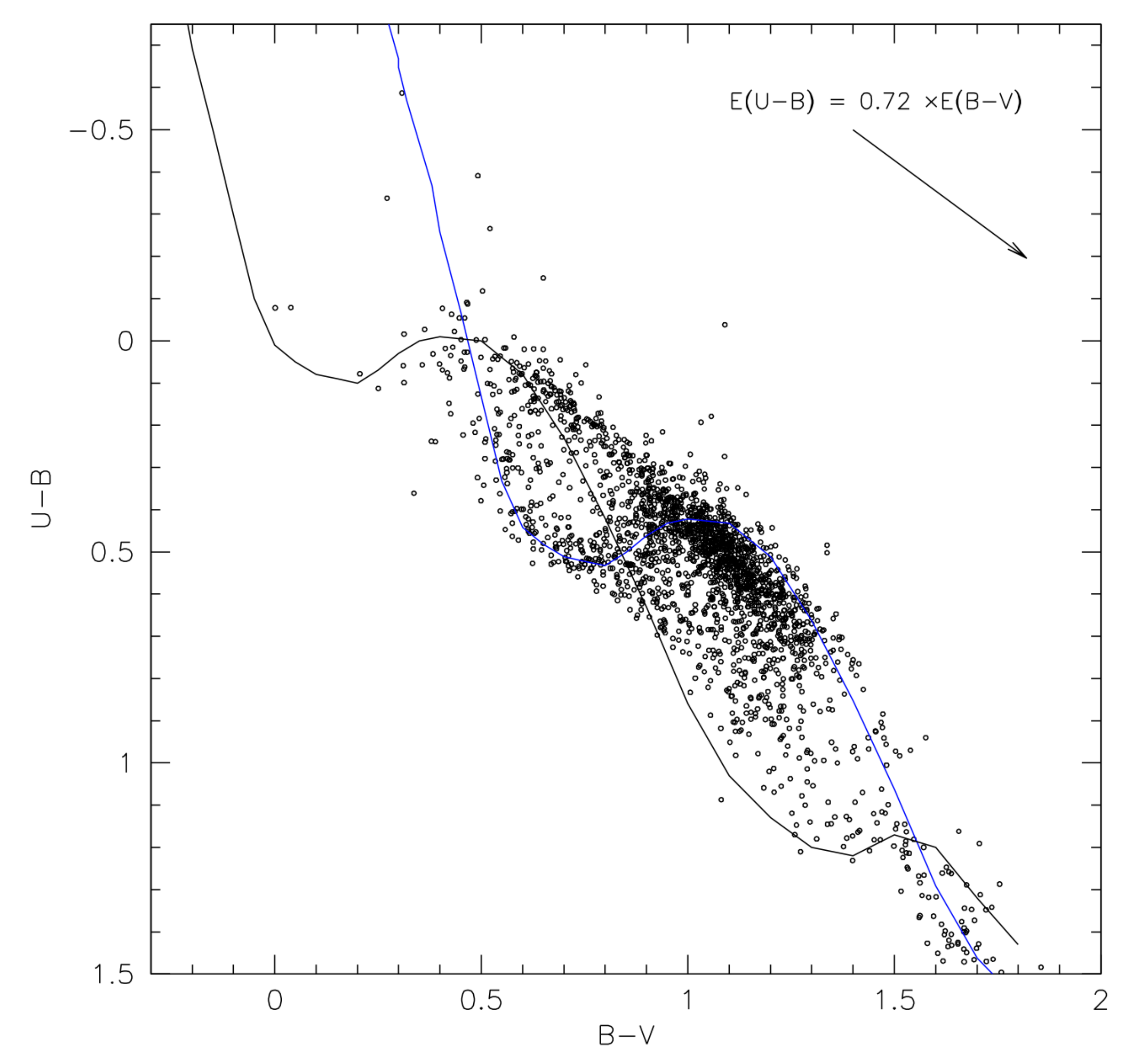}
\caption{TCD diagram of Hogg~19. Only stars having $\sigma_{U,B,V}\leq0.02$ are shown. See text
for details.}
\end{figure}

\begin{figure}
\includegraphics[width=\columnwidth]{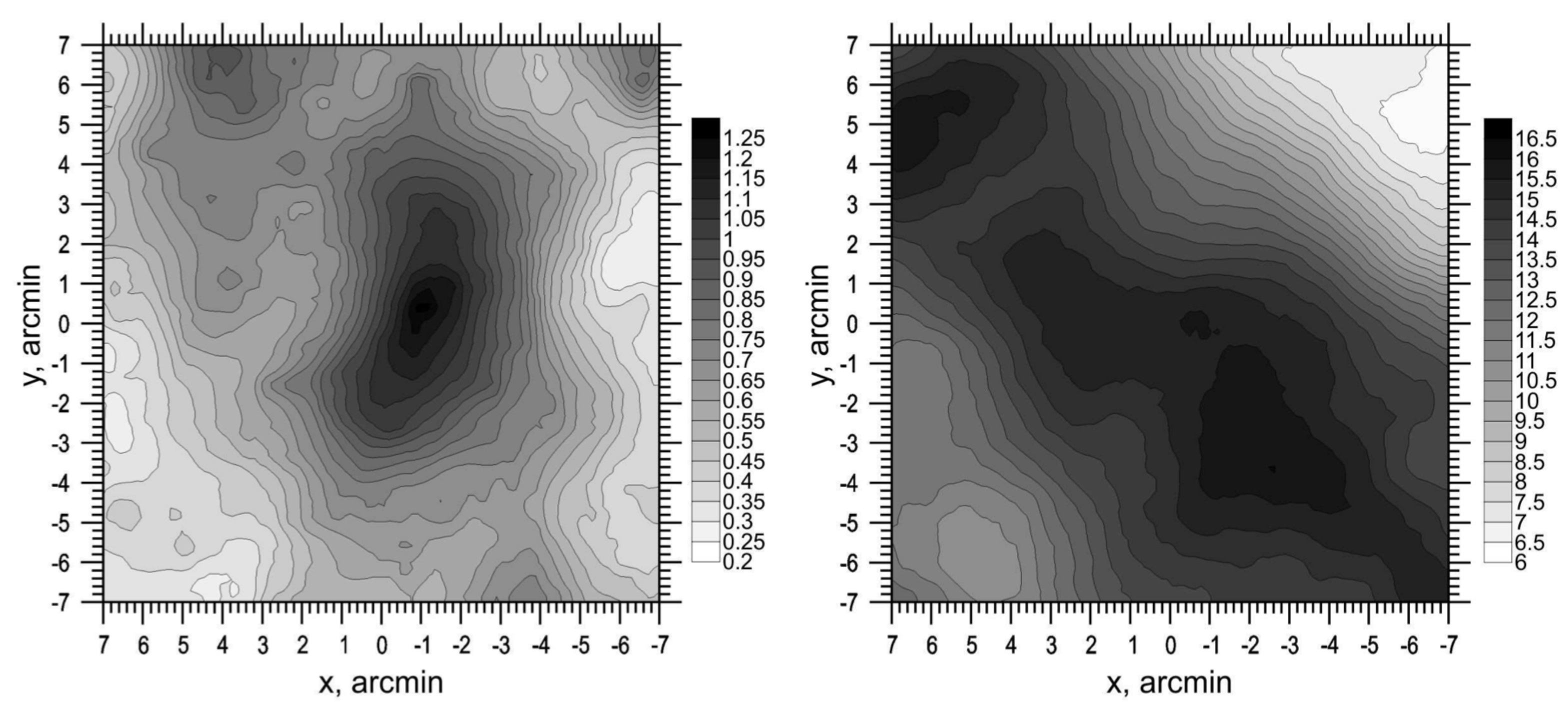}
\caption{Surface density map for stars bright than $V=14$, which sample Hogg~19 according to
Kharchenko et al. (2013) solution (left panel), and for stars fainter than $V=18$, to reproduce our
solution.}
\end{figure}

\begin{figure}
\includegraphics[width=\columnwidth]{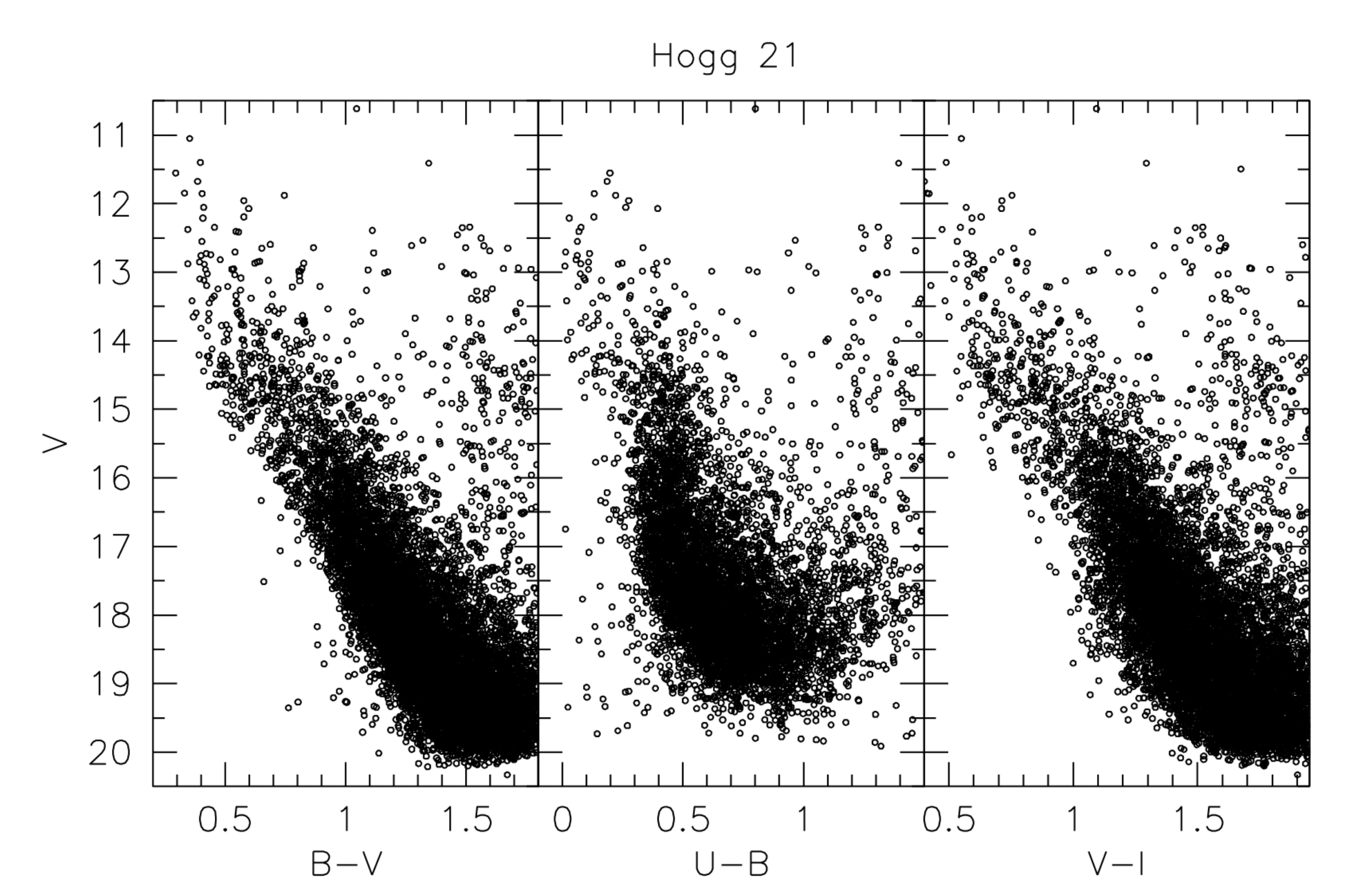}
\caption{CMDs of Hogg~21 in $V/B-V$ (left panel), $V/U-B$ (middle panel), and $V/V-I$ (right panel).}
\end{figure}

\begin{figure}
\plottwo{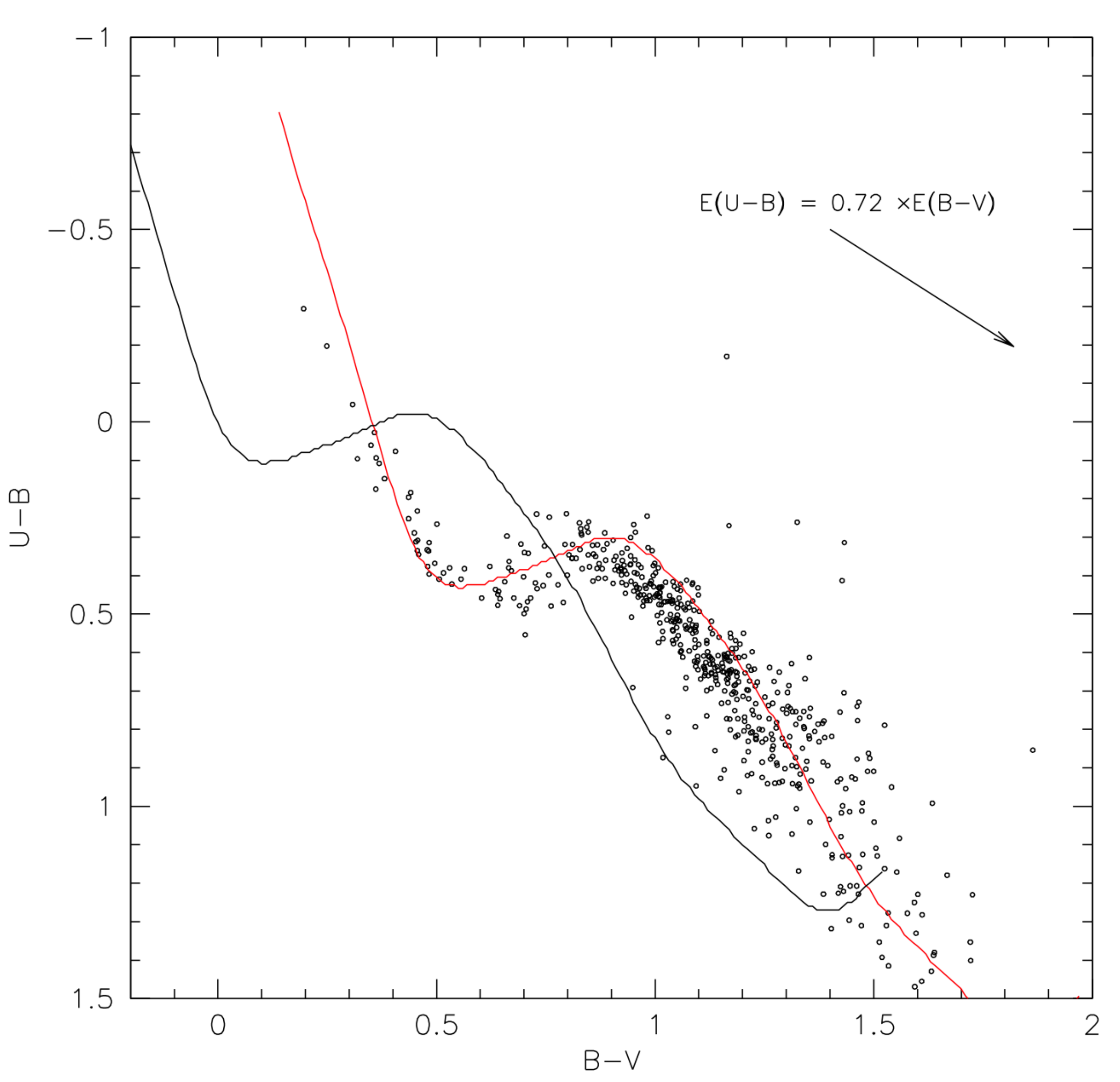}{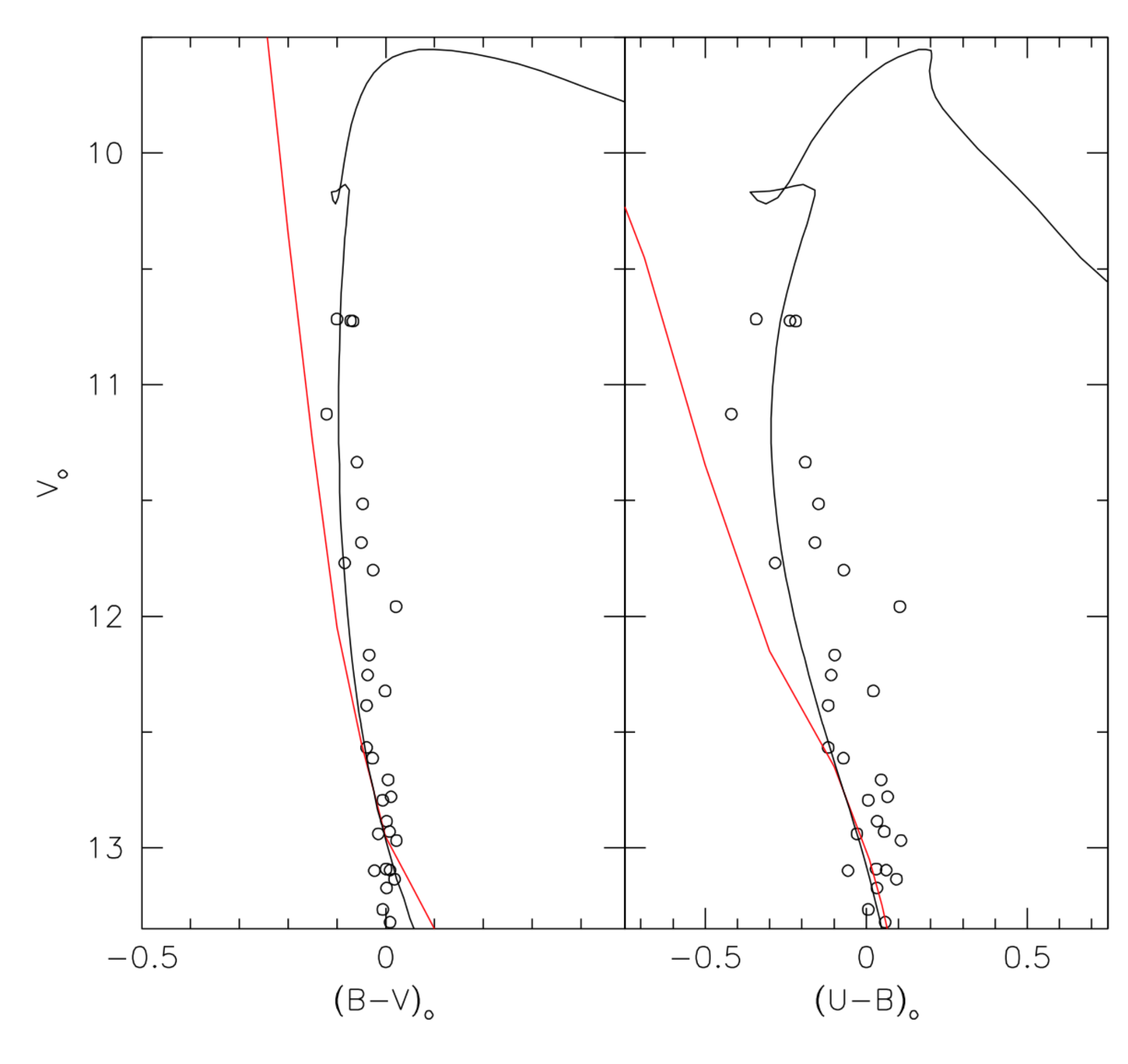}
\caption{Left panel: TCD of Hogg~21, and reddening solution. Right panel: distance and age solution.
See text for details,}
\end{figure}

\begin{figure}
\includegraphics[width=\columnwidth]{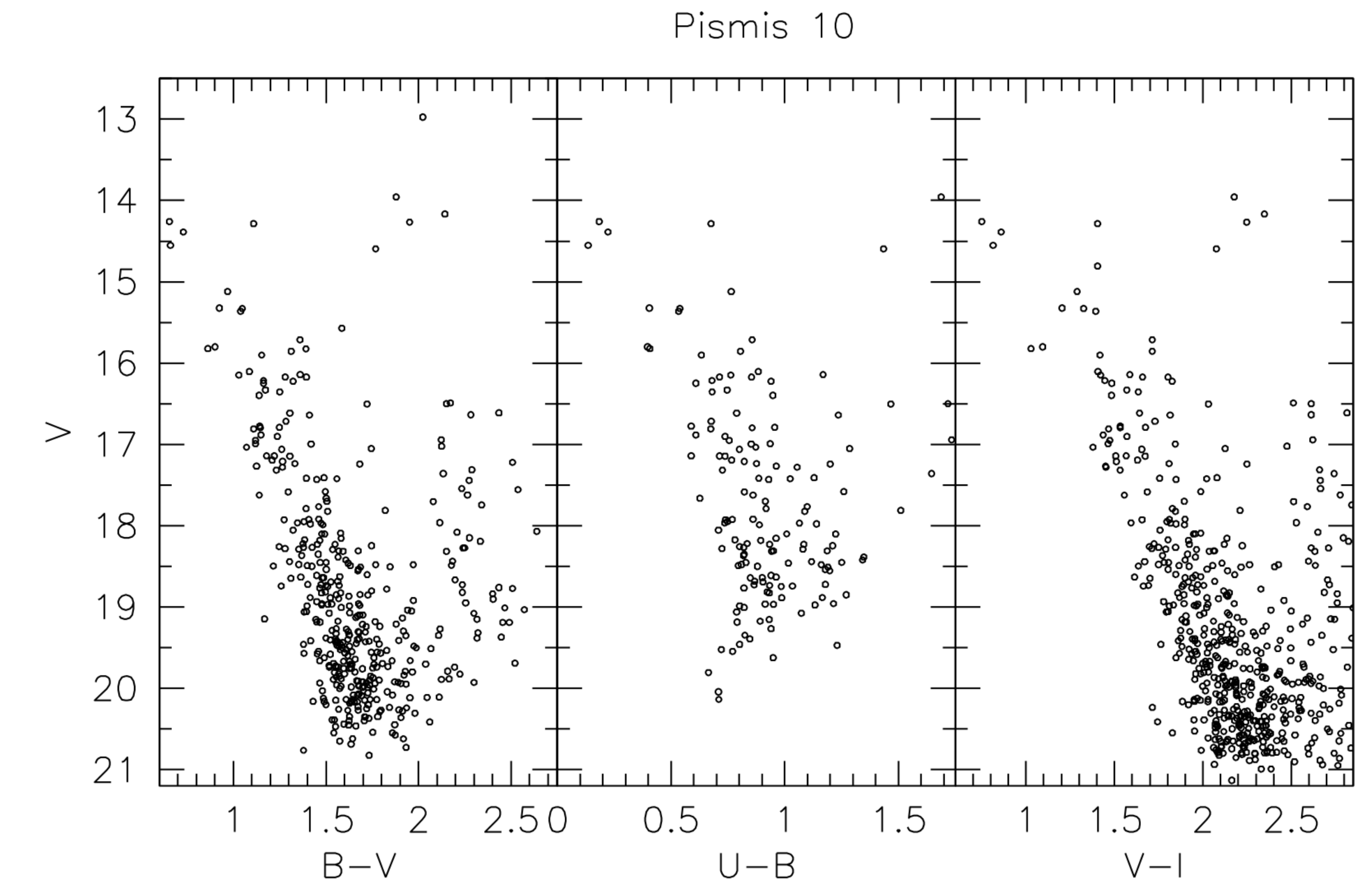}
\caption{CMDs of Pismis~10 in $V/B-V$ (left panel), $V/U-B$ (middle panel), and $V/V-I$ (right panel).}
\end{figure}

\begin{figure}
\plottwo{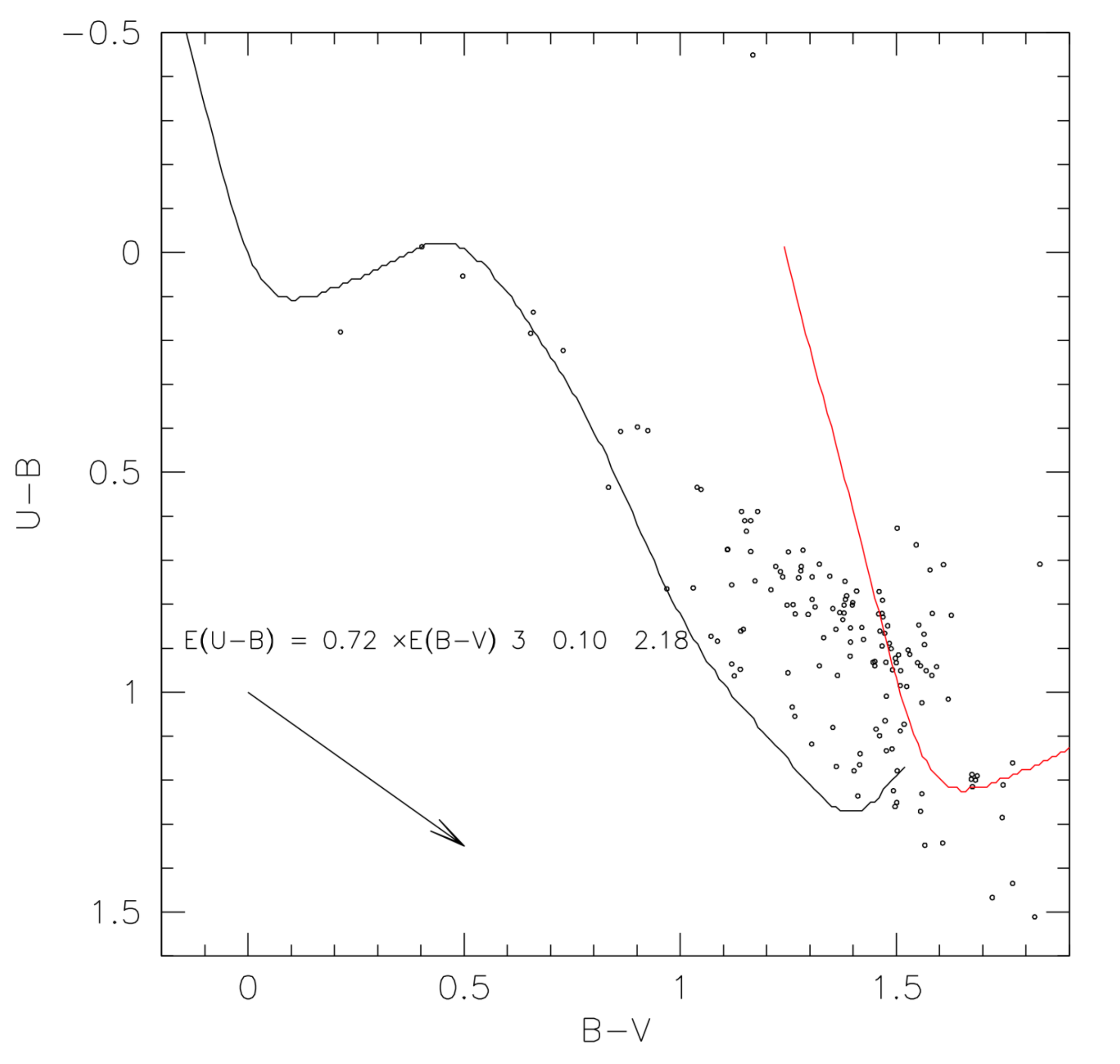}{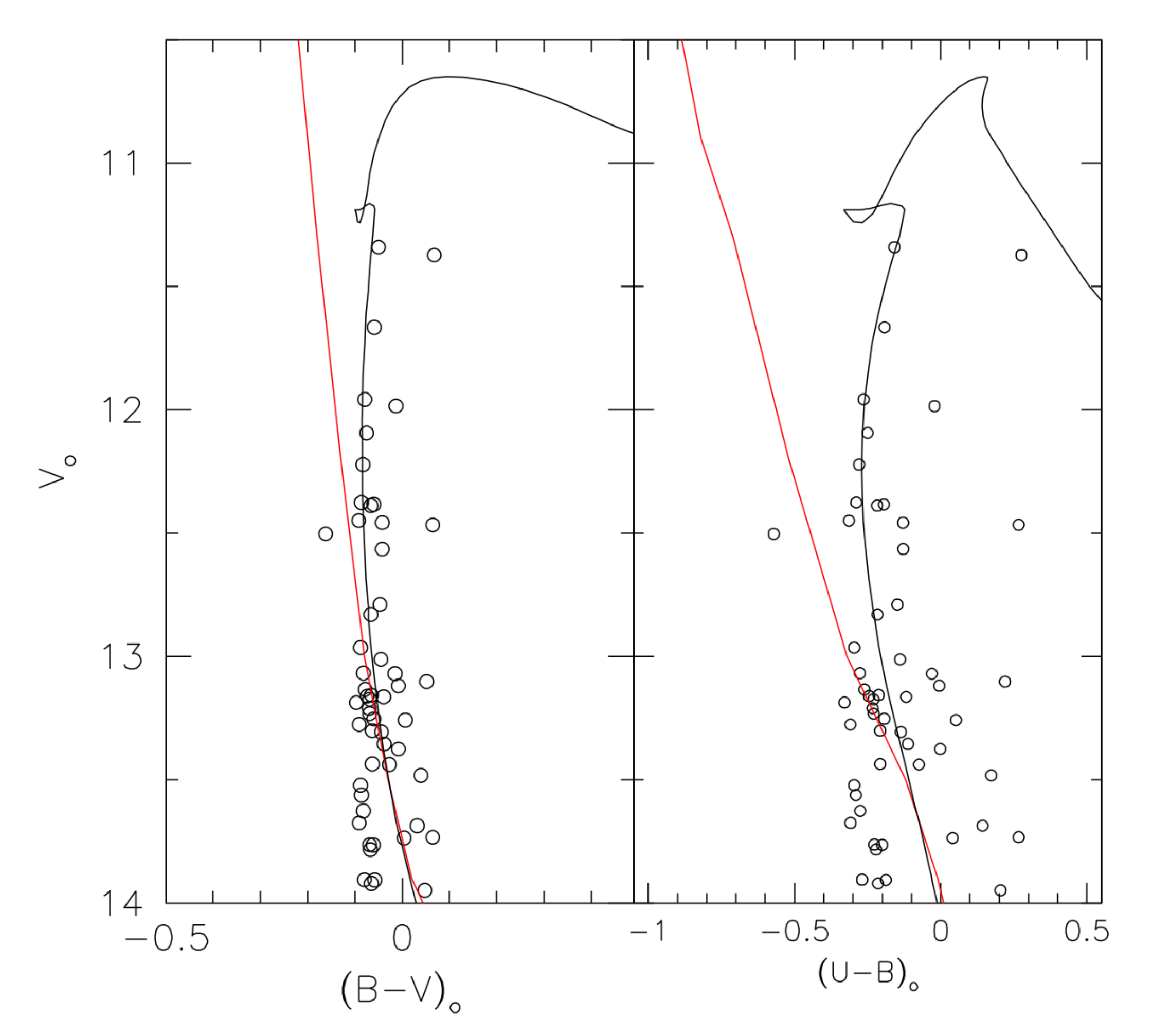}
\caption{Left panel: TCD of Pismis~10, and reddening solution. Right panel: distance and age solution.
See text for details.}
\end{figure}

\begin{figure}
\includegraphics[width=\columnwidth]{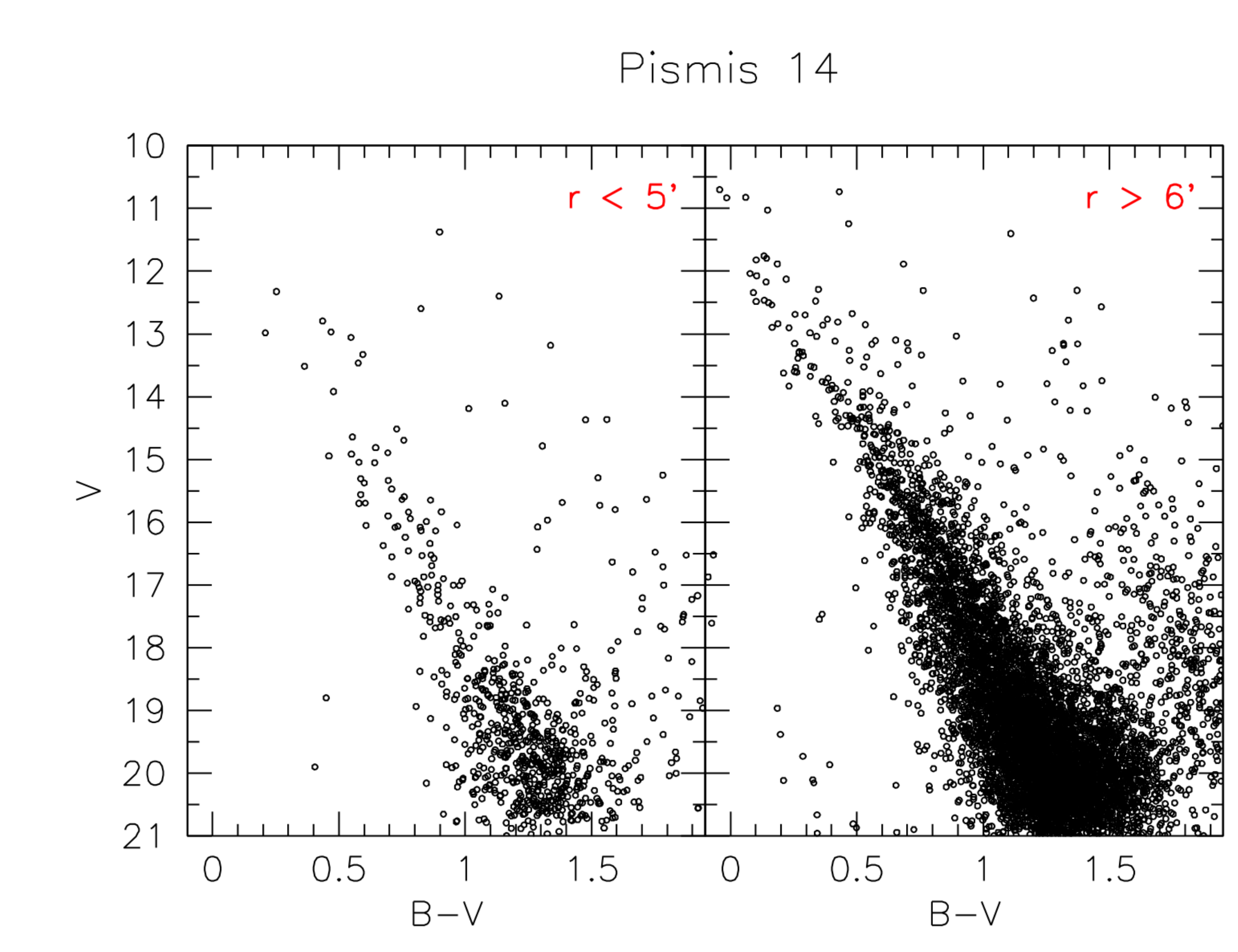}
\caption{CMDs of Pismis 14 in $V/B-V$.  Left panel: stars within the cluster estimated radius to
exclude contamination from NGC~2910. Right panel: stars outside the cluster area, to highlight NGC~2910.}
\end{figure}

\begin{figure}
\includegraphics[scale=0.7]{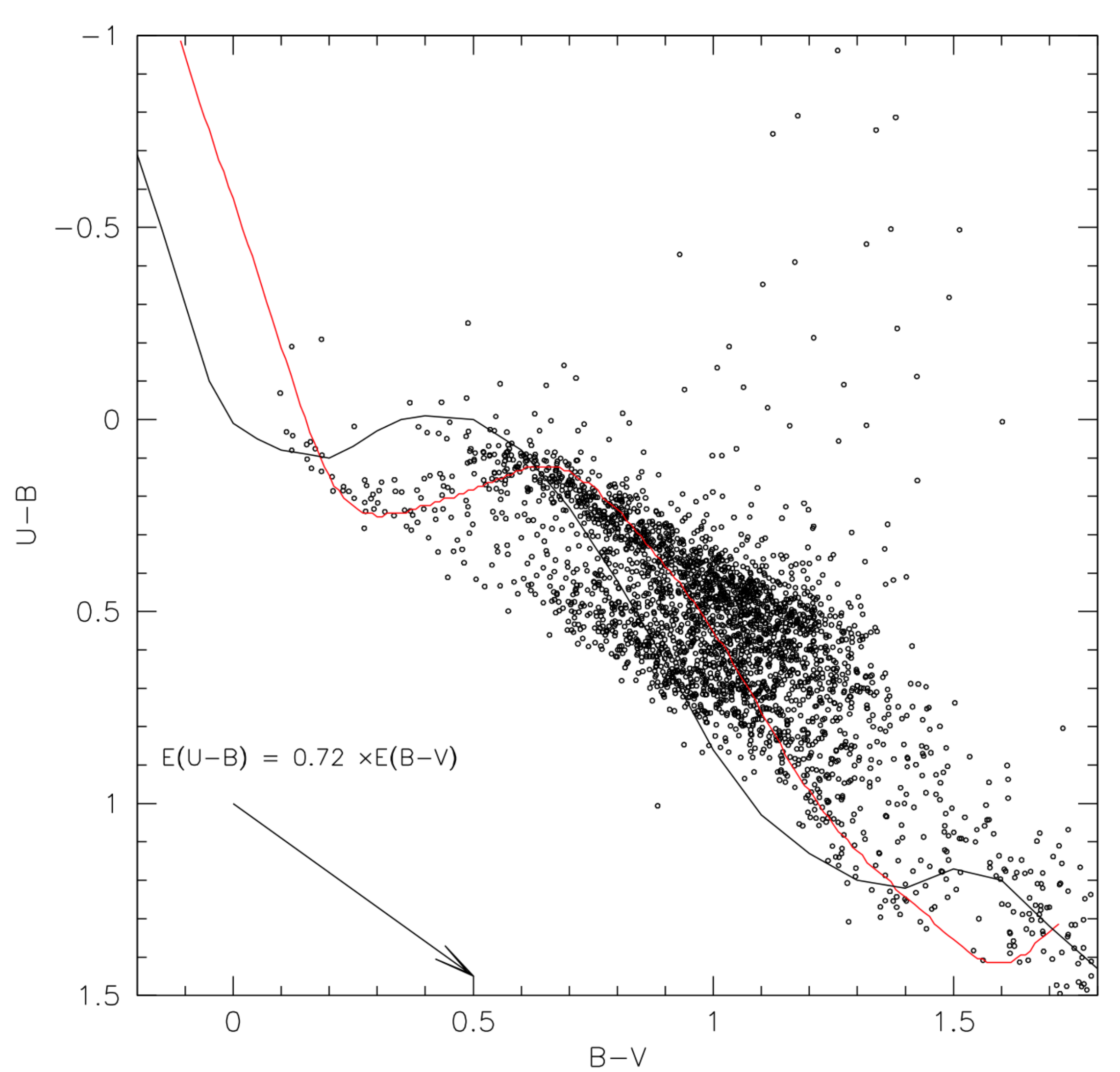}
\caption{TCD of Pismis~14 and NGC~2910, and reddening solution for NGC~2910.}
\end{figure}
\clearpage

\begin{table}
\tabcolsep 0.1truecm
\caption{Stellar fields studied in this work}
\begin{tabular}{lrrcr}
\hline
Cluster & RA(2000.0) & Dec (2000.0) &  $l$ & $b$ \\
\hline
   & [deg]  & [deg] & [deg] & [deg] \\
\hline\hline
Pismis~10    &   135.6500 &  -42.6330 & 265.429  & 1.960 \\
Pismis~14    &   142.4625 &  -52.7833  & 275.150  & -1.145 \\
Trumpler~22  &  217.7583&  -61.1667 & 314.647  &  -0.581\\
Lynga~6      & 241.2167  &  -51.9333 & 330.369 &  0.323\\
Hogg~19      &  247.2375 &  -49.1000 &  335.088  &   -0.302\\
Hogg~21      &  251.4041 &  -47.7333 &  337.956  &   -1.437\\
\hline\hline
\end{tabular}
\end{table}

\clearpage

\begin{table}
\tabcolsep 0.02truecm
\caption{Log of UBVI photometric observations for the fields under study.}
\begin{tabular}{lcccc}
\hline
\noalign{\smallskip}
Date & Field & Filter & Exposures (sec) & airmass (X)\\
\noalign{\smallskip}
\hline\hline
\noalign{\smallskip}
2006 Mar 19    & Trumpler~22 & $U$ & 10, 30, 200, 1800 & 1.17$-$1.24\\
                     &                        & $B$ & 7, 30, 100, 900 & 1.01$-$1.02\\
                      &                       & $V$ & 5, 30,100, 700 & 1.02$-$1.10 \\
                       &                      & $I$  & 5, 10, 30,100,600 & 1.02$-$1.11 \\
2009 Mar 19 & Hogg 19 & $U$ & 30,2x200,2000     & 1.17$-$1.19 \\
            &            & $B$ & 2x20, 150,1500      & 1.30$-$1.33 \\
                         &                         & $V$& 10,100,900& 1.17$-$1.19\\
                          &                        & $I$& 10,100,900& 1.39$-$1.42\\
2009  Mar 21 & Pismis 14 &  $U$ & 10, 30, 200, 1800 & 1.17$-$1.24\\
                     &                        & $B$ & 7, 30, 100, 900 & 1.01$-$1.02\\
                     &            & $V$ & 3x10,900     & 1.6$-$1.17 \\
                    &            & $I$ & 10,100,900   & 1.21$-$1.23 \\
                    & Lynga 6 & $U$ & 10, 30, 200, 1800 & 1.17$-$1.24\\
                    &                        & $B$ & 7, 30, 100, 900 & 1.01$-$1.02\\
                    &    & $V$ & 10,100,900 & 1.02$-$1.04 \\
                   &     & $I$ & 10.100.900 & 1.01$-$1.02\\
2009  Mar 22 &  Pismis 10  & $V$ & 20,100,900   & 1.15$-$1.16 \\
            &            & $B$ & 20,150, 1500     & 1.05$-$1.06 \\
            &            & $U$ & 200,2000       & 1.10$-$1.02\\
            &            & $I$ & 20, 100,900   & 1.10$-$1.11 \\
             & Hogg 10  & $U$ & 20,100,900   & 1.15$-$1.16 \\
            &            & $B$ & 20,150, 1500     & 1.05$-$1.06 \\
                   &  & $V$ & 10,100,900 & 1.02$-$1.05 \\
                   &     & $I$ & 10.100.900 & 1.01$-$1.03\\
            \hline
\noalign{\smallskip}
\hline
\end{tabular}
\end{table}

\clearpage

\begin{table*}
\tabcolsep 0.4truecm
\caption{Night by night photometric solutions.}
\begin{tabular}{ccccc}
\hline\hline
Night  &    19/03/2006  &    19/03/2009 &     21/03/2009  &        22/03/2009  \\
\hline\hline
u1    & -0.747$\pm$0.003   &-0.884$\pm$0.006& -0.860 $\pm$0.008  &  -0.879$\pm$0.007 \\
u2    &  0.45                        &     &                                 &                                  \\
u3    & -0.015$\pm$0.006   &-0.037$\pm$0.009&-0.017 $\pm$0.012   & -0.023$\pm$0.010   \\
rms  &  0.02                        &0.09 &   0.10                       &  0.09                         \\
\hline
b1             &-1.980$\pm$0.013    &-2.085$\pm$0.010 &   -2.063$\pm$0.010 &   -2.068$\pm$0.010 \\
b2             & 0.25                         &                               &                               &                                         \\
b3             & 0.153 $\pm$0.017    &0.150$\pm$0.010 & 0.128$\pm$0.010    &  0.132$\pm$0.010  \\
rms           & 0.03                          &       0.08                & 0.05                       &         0.08                 \\
\hline
v1$_{bv}$  & -2.199$\pm$0.009   &-2.136$\pm$0.028   &   -2.126$\pm$0.006 &   -2.127$\pm$0.006  \\
v2$_{bv}$  &  0.16                         &                                &                                 &                                 \\
v3$_{bv}$  &-0.060$\pm$0.012    & -0.028$\pm$0.005 &    -0.035$\pm$0.006 &    -0.036$\pm$0.006 \\
rms            &  0.02                        &      0.04                     & 0.05                           &    0.05                     \\
\hline
v1$_{vi}$ & -2.198$\pm$ 0.005 &-2.148$\pm$0.005&    -2.121$\pm$0.006 &    -2.124$\pm$ 0.006  \\
v2$_{vi}$  & 0.16                     & & \\
v3$_{vi}$ &-0.063$\pm$0.005 &-0.013$\pm$0.045&   -0.038$\pm$0.005 &    -0.032$\pm$0.005 \\
rms   & 0.03                             &  0.05&   0.05       &        0.04        \\
\hline
i1   &  -1.298$\pm$0.009  &  -1.321$\pm$0.004& -1.319$\pm$0.005  &  -1.313$\pm$0.005  \\
i2    &   0.08                      &                                &  &                                                               \\
i3    & -0.054$\pm$0.009  & -1.010$\pm$0.003& -0.017$\pm$0.004   & -0.014$\pm$0.003   \\
rms &   0.02                      & 0.04 & 0.04                       &   0.04                                            \\
\hline\hline
\end{tabular}
\end{table*}

\clearpage

 \begin{table}
\tabcolsep 0.1truecm
\caption{Basic parameters for our cluster sample}
\begin{tabular}{rccccccc}
\hline
Cluster & $V_{lim}$& RA(2000.0) & Dec (2000.0) &  $R$  & E(B-V) & d$_{\odot}$ & Age\\
\hline
   & mag  & [deg] & [deg] & armin & & kpc & Myr\\
\hline\hline
Trumpler~22  & 18 &  217.76005 &  -61.17384   &    6.4$\pm$0.5  &0.48$\pm$0.05 & 1.9$\pm$0.1 &70$\pm$10 \\
Lynga~6         & 16 &  241.16964 &  -51.94483  &    5.5$\pm$0.3  &1.20$\pm$0.10 & 2.0$\pm$0.2 & 79$\pm$10\\
Hogg~19        & 19 &  247.21001 &  -49.13633  &   $\geq$7          &0.80$\pm$0.2     & 2.6$\pm$0.5 &2500$\pm$200\\
Hogg~21        & 14 &  251.33181 &  -47.72750  &    3.9$\pm$0.3  &0.48$\pm$0.10  & 2.1$\pm$0.2 & 100$\pm$10\\
Pismis~10      & 18 &  135.65230 &  -43.64383  &    6.1$\pm$0.3  & 1.5$\pm$0.10  & 2.7$\pm$0.3 & 250$\pm$20\\
Pismis~14      & 18 &  142.51458 &  -52.72183  &     $\geq$5        &&  &\\
\hline\hline
\end{tabular}
\end{table}

\end{document}